\documentclass[prb,twocolumn,showpacs,floatfix,amsmath,amssymb,mathptm,bm]{revtex4-1}
\usepackage{graphicx}
\usepackage{epstopdf}
\usepackage{wasysym}
\usepackage[dvips]{color}
\usepackage{hyperref}
\usepackage{hypernat}
\usepackage{afterpage}
\usepackage{sidecap}
\usepackage{float}
\usepackage{footnote}
\usepackage{threeparttable}
\def\etal{{\em{et al.}}}
\newcommand{\longoverbrace}[2]{\overbrace{#1}^{\text{\hbox to 0cm{\hss #2 \hss}}}}  
\newcommand{\longunderbrace}[2]{\underbrace{#1}_{\text{\hbox to 0cm{\hss #2 \hss}}}} 
\DeclareRobustCommand{\specialp}{\ensuremath{\mathcal{P}}}
\def\K{\mathbf{K}}
\def\R{\mathbf{R}} 
\def\Q{\mathbf{Q}}

\begin{document}
\title{Finite Cluster Typical Medium Theory for Disordered Electronic Systems}
\author{C. E. Ekuma}
\altaffiliation{Electronic address: cekuma1@lsu.edu}
\affiliation{Department of Physics \& Astronomy, Louisiana State University,
Baton Rouge, Louisiana 70803, USA}
\affiliation{Center for Computation \& Technology, Louisiana State University, Baton Rouge, Louisiana 70803, USA}

\author{C. Moore}
\affiliation{Department of Physics \& Astronomy, Louisiana State University,
Baton Rouge, Louisiana 70803, USA}
\affiliation{Center for Computation \& Technology, Louisiana State University, Baton Rouge, Louisiana 70803, USA}

\author{H. Terletska}
\affiliation{Department of Physics \& Astronomy, Louisiana State University,
Baton Rouge, Louisiana 70803, USA}
\affiliation{Center for Computation \& Technology, Louisiana State University, Baton Rouge, Louisiana 70803, USA}

\author{K.-M. Tam}
\affiliation{Department of Physics \& Astronomy, Louisiana State University,
Baton Rouge, Louisiana 70803, USA}
\affiliation{Center for Computation \& Technology, Louisiana State University, Baton Rouge, Louisiana 70803, USA}

\author{N. S. Vidhyadhiraja}
\affiliation{Theoretical Sciences Unit, Jawaharlal Nehru Center for Advanced Scientific Research, Bangalore, 560064, India}

\author{J. Moreno}
\affiliation{Department of Physics \& Astronomy, Louisiana State University,
Baton Rouge, Louisiana 70803, USA}
\affiliation{Center for Computation \& Technology, Louisiana State University, Baton Rouge, Louisiana 70803, USA}

\author{M. Jarrell}
\altaffiliation{Electronic address: jarrellphysics@gmail.com}
\affiliation{Department of Physics \& Astronomy, Louisiana State University,
Baton Rouge, Louisiana 70803, USA}
\affiliation{Center for Computation \& Technology, Louisiana State University, Baton Rouge, Louisiana 70803, USA}

\relpenalty=10000      
\binoppenalty=10000

\begin{abstract}
\noindent We use the recently developed typical medium dynamical cluster (TMDCA) 
approach~[Ekuma \etal,~\textit{Phys. Rev. B \textbf{89}, 081107 (2014)}] 
to perform a detailed study of the Anderson localization transition in three dimensions for the
Box, Gaussian, Lorentzian, and Binary disorder distributions, and 
benchmark them with exact numerical results. Utilizing the nonlocal 
hybridization function and the momentum resolved typical spectra to characterize the 
localization transition in three dimensions, we demonstrate the importance of both spatial 
correlations and a typical environment for the proper characterization of the localization 
transition in all the disorder distributions studied. As a function of increasing cluster size, 
the TMDCA systematically recovers the re-entrance behavior of the mobility edge for disorder 
distributions with finite variance, obtaining the correct critical disorder strengths, and 
shows that the order parameter critical exponent for the Anderson localization 
transition is universal. The TMDCA is computationally efficient, requiring only a small cluster to 
obtain qualitative and quantitative data in good agreement with numerical exact results at a 
fraction of the computational cost. Our results demonstrate that the TMDCA provides 
a consistent and systematic description of the Anderson localization transition. 
\end{abstract}


\maketitle
\section{Introduction}
\label{sec:intro}
Disorder is ubiquitous in materials and can drastically alter their properties, in particular, 
their electronic structure and transport properties, and even may induce electron localization. This 
phenomenon is known as the Anderson metal-insulator-transition.~\cite{PhysRevB.89.081107,Anderson,gang4,Kramer,Mirlin-RMP} 
Here, the transition from a metal to an insulator is not characterized by the vanishing of 
the charge carrier density but by the cancellation of the hybridization paths accompanying the quantum 
localization of the wave functions due to coherent backscattering from random impurities, deep-trapped 
states, etc.  As a result, electrons that occupy such exponentially localized states are restricted to 
finite regions of space, and hence cannot contribute to transport.  The Anderson insulator is gapless 
indicating that the single-particle excitations are essential in determining its physical 
properties especially at low energies.  While there has been significant progress in the quest 
to understand this phenomenon, a proper effective mean-field treatment is not yet fully 
developed. 

There have been numerous theoretical studies of disordered electron systems employing computational techniques 
of varying complexity not limited to numerically exact methods including exact diagonalization, 
transfer matrix, and kernel polynomial methods,\cite{SILVER1994,PhysRevE.56.4822,KPM_review_2006,
Schubert,PhysRevLett.47.1546,Kramer,Kramer1996,MacKinnonKramer1983,Kramer2010,Markos-review-2006,Slevin2014} 
various renormalization group techniques,~\cite{PhysRevB.20.4726,PhysRevB.33.7738,0022-3719-18-10-015} 
and mean-field theories.~\cite{Abou-Chacra,Vollhardt_SCT,PhysRevB.81.155106,Vollhardt-review,PhysRev.156.809,Velicky68,Vlad2003}
While numerical exact methods have been successfully used to study Anderson
localization, they generally require the treatment of large clusters and the use of
powerful computers.  As a result, they are difficult to extend to the treatment of 
interacting systems or chemically specific models.   
An alternative approach is offered by mean-field theories such as the coherent 
potential approximation and its extensions.~\cite{PhysRev.156.809,Velicky68,Jarrell01} 
They map the lattice onto relatively small self-consistently
embedded clusters.  These methods have been successfully extended to the treatment of 
interacting disordered systems and to chemically realistic models.  Unfortunately,
these methods have been woeful in the treatment of the Anderson localization due 
mainly to the averaging procedure utilized and improvements in the environment describing the 
effective medium have been limited to single-sites. 

Due to the central role a mean-field theory that properly describes the Anderson localization transition 
(ALT) may play for further progress in the study of electron localization, there is a need to formulate 
such theory. Furthermore,  
a well-known long historical dichotomy exists between the mean-field results and the numerical 
data for the Anderson localization transition. Here, we demonstrate that the dichotomy can be 
reconciled by incorporating spatial fluctuations in a typical environment into the mean-field 
theory thereby offering a solution to this long-standing problem and providing a systematic framework 
in the mean-field theory of the Anderson localization. 

The most commonly used self-consistent mean-field theory, in the study of disordered electron 
systems, is the coherent  potential approximation (CPA).~\cite{PhysRev.156.809,Velicky68} In 
the CPA, the original disordered lattice is replaced by an impurity placed in an averaged 
local (momentum-independent) effective medium.  While the CPA successfully describes 
some one-particle properties, such as the density of states (DOS) in substitutional disordered 
alloys,~\cite{PhysRev.156.809, Kirpatrick-Velickyprb70} it fails to describe the Anderson 
localization transition. This failure stems, in part, from the single-site nature of the 
CPA, as it is unable to capture crucial multiple backscattering interference effects that lead 
to electron localization. Cluster extensions of the CPA, like the dynamical cluster approximation 
(DCA)~\cite{PhysRevB.63.125102,PhysRevB.61.12739,Jarrell01} and molecular CPA~\cite{MCPA}, allow 
for the incorporation of such nonlocal spatial correlations; however, they still fail to describe 
the Anderson localization transition. The arithmetic averages of random one-particle quantities 
(e.g., density of states) calculated within such mean-field theories cannot distinguish between 
extended and localized states and are not critical at the Anderson 
transition.~\cite{Thouless1974,Thouless1970,PhysRevB.89.081107,0953-8984-26-27-274209} Hence, such 
average quantities cannot be used as an order parameter. This is the reason that most mean-field 
theories like the CPA \cite{RevModPhys.46.465} and its cluster extensions including the 
DCA,~\cite{PhysRevB.63.125102,RevModPhys.77.1027,PhysRevB.61.12739} fail to provide a proper description 
of the Anderson localization in disordered systems. This failure is intrinsic to these theories as the 
algebraically averaged quantities, i.e., the averaged density of states, always favor the metallic 
state. This can be understood from the fact that in an infinite system of localized states, 
the average density of states is nothing but the global density of states, which is a smooth function of 
the disorder strength near the critical point while the local density of states becomes discrete with a 
non-trivial system size dependence (see, e.g., Refs.\onlinecite{nakayama2003fractal,Thouless1974,
Thouless1970,PhysRevLett.72.526,Lagendijk2009,
Janssen1998, PhysRevB.74.153103, Vlad2003} for a detailed discussion). Further, 
due to the large fluctuations in the local Green function, 
its \textit{typical value} is far removed from the average one~\cite{Miranda2012} as such, the local average Green 
function, which characterizes these mean-field environments, does not have any information about the 
\textit{typical nature} of the local order parameter needed to signal a phase transition. 
In Sec.~\ref{sec:absence_DCA}, we will demonstrate the failures of finite cluster extensions of 
the CPA in characterizing the Anderson localization transition using the DCA. 

Finding a proper single-particle order parameter for the Anderson localization transition capable of 
distinguishing between the localized and extended states is a major challenge in the study of disordered 
electronic systems.  In contrast to the arithmetic average, the geometrical average
\cite{Vlad2003,Janssen1998,Janssen1994,PhysRevLett.72.526,Crow1988} gives a better approximation 
to the most probable value of the local density of states. Dobrosavljevi\'{c} \etal~\cite{Vlad2003} 
developed the typical medium theory (TMT) to study disordered systems, where the typical density 
of states (TDOS), approximated using the geometrical averaging over disorder configurations, is used 
instead of the arithmetically averaged local density of states.  They demonstrated that the TDOS 
vanishes continuously as the strength of the disorder increases towards the critical point and it 
can be used as an effective mean-field order parameter for the Anderson localization transition. 

While the single-site TMT has been shown to be successful in describing localized electron states, 
it still has several drawbacks. In particular, it does not provide a proper description of the critical 
behavior of the Anderson localization transition in three dimensions (3D).  As a local approximation, the TMT neglects the
effects of coherent backscattering and, as a result, the critical disorder strength obtained is 
$W_c^{TMT}=1.65$ instead of the numerically exact value 
$W_c\approx2.1$~\cite{Slevin99,PhysRevB.63.045108,Mirlin-RMP,PhysRevB.84.134209,PhysRevLett.47.1546,
Bulka87,Kramer1987,PhysRevB.51.663} 
for the box distribution (in our units). Also, the \textit{universal} order 
parameter critical exponent (which has also been called the \textit{typical order parameter 
exponent})~\cite{Janssen1994} $\beta$ obtained in the local TMT is $\beta^{TMT}\approx1.0$ whereas
its recently reported value is $\beta\approx1.67$.~\cite{PhysRevLett.105.046403,PhysRevB.84.134209}  
Another crucial drawback of the single-site TMT in 3D is its inability to capture the 
re-entrance behavior of the mobility edge (energy separating extended and localized electron states)
demonstrated in exact numerical studies~\cite{Bulka85,Bulka87,Fehske} 
for the disorder distributions with finite variance: the box and Gaussian disorder distributions. 
The TMT also underestimates the 
extended state regions in all the disorder distributions studied in this paper.

The inadequacies of the single-site TMT can be remedied by incorporating spatial 
(nonlocal) correlations by constructing its cluster extension. This can be achieved by using the 
DCA or molecular CPA schemes, which systematically incorporate the missing nonlocal effects.  

In this paper, we show in detail a successful extension of the local TMT to its cluster version 
using ideas from the DCA. We demonstrate how the finite cluster extension of the 
local TMT is able to systematically solve all the crucial drawbacks of the single-site TMT, 
indicating the necessity to include the missing nonlocal effects. 
One of the features of the Anderson localization transition in three dimensions 
is the non-self-averaging nature of the local quantities close to the localization 
transition, which obtain a highly skewed (log-normal) distribution.  
Hence special care must be taken in constructing a cluster extension of the TMT. 
To avoid such self-averaging issues in the TDOS, we developed the typical medium dynamical cluster 
approximation (TMDCA),~\cite{PhysRevB.89.081107}, which properly characterizes the Anderson localization 
transition in 3D and does not suffer from the self-averaging, by 
explicitly separating the local part of the TDOS and treating it with a geometric average over disorder 
configurations. Hence, we are able to obtain a proper TDOS that 
characterizes the Anderson localization in 3D. We demonstrate the versatility of our method by applying it to 
the box, Gaussian, Lorentzian, and binary disorder distributions
and benchmark it with numerically exact methods.

The typical medium dynamical cluster approximation scheme is demonstrated to be a systematic, 
self-consistent effective medium theory for characterizing electron localization. As a function of 
increasing cluster size, we demonstrate that the TMDCA achieves convergence of both the critical 
disorder strength and the trajectories of the mobility edge as a function of the cluster size. Furthermore, 
the TMDCA fulfills all the essential requirements expected of a ``successful'' cluster 
theory.~\cite{a_gonis_92,PhysRevB.63.125102}  We find that the TMDCA scheme is
a systematic self-consistent effective medium theory to study Anderson localization transition in 
three-dimensions, which i) recovers the original single-site TMT scheme at cluster size $N_c$ = 1;  
ii) recovers the DCA results at small disorder strength (when most states are metallic); iii) provides 
a proper way to separate the energy scales such that the characteristic mobility edge behavior 
(for the disorder distributions with finite variance) is recovered; iv) captures the critical behavior 
of the Anderson localization transition with correct critical disorder strength  $W_c$ and order parameter 
critical exponent $\beta$, and provides the correct description of the Anderson insulator at large 
disorder strength (when all states are localized); and v) fulfills all the essential requirements expected 
of a ``successful'' cluster theory.~\cite{a_gonis_92,PhysRevB.63.125102} 

The main problem addressed in this paper is how the mobility edge energies vary with disorder strength, their 
trajectories, and what happens to these trajectories in the proximity of the Anderson localization 
transition. Furthermore, since the DCA always becomes exact when $N_c \rightarrow \infty$, the main role of the effective medium 
in approaches based on the DCA, or its extensions, is to accelerate this convergence. For the ALT, we find that 
the effective medium formed from the average Green function does not converge as $N_c$ becomes large, rather 
it is only able to describe the precursors to localization (cf. Section~\ref{sec:absence_DCA}). However, we find 
that a number of effective media based upon the geometrically averaged density of states provide convergent results, 
i.e., an order parameter. So far, we find that the fastest convergence is provided by the TMDCA.

The rest of this paper is organized as follows: following the introduction in this Section~\ref{sec:intro}, 
we present the model and describe the details of the formalism in Section~\ref{sec:formalism}. We present 
the results of our calculations in Section~\ref{sec:results}. The absence of localization in the DCA (shows 
only precursor to localization) is described in~\ref{sec:absence_DCA}.  Detailed 
analysis of how we avoid self-averaging as the size of the cluster is increased is discussed in~\ref{sec:self-averaging}.
Then in~\ref{sec:pole_procedure}, we describe how to treat the states close to criticality where the hybridization 
function vanishes leading to the development of poles (delta functions) in the imaginary part of 
the cluster-excluded Green function, ${\cal G}(K,\omega)$.  
A detailed analysis of our results for the box disorder distribution is presented in Sec.~\ref{sec:boxdistribution}. 
Section~\ref{sec:binarydistribution} presents our results for the binary disorder distribution. In 
Sec.~\ref{sec:Gaussiandistribution}, we present our results for the Gaussian disorder distribution while 
Sec.~\ref{sec:Lorentziandistribution} shows the results of our computations for the Lorentzian distribution. 
In Sec.~\ref{criticalparameters}, we discuss in detail the procedure for obtaining the critical parameters 
especially the order parameter critical exponent $\beta$ for the various disorder distributions,
while in Sec.~\ref{sec:difficultyinME}, we address the discrepancy observed in the 
trajectories of the mobility edge energies at higher disorder strength.
We summarize and discuss future directions in Sec.~\ref{sec:conclusion}. In appendices~\ref{TMM_appendix} and 
\ref{KPM_appendix}, we present a concise description of the developed transfer matrix and kernel polynomial methods, 
respectively, used in benchmarking the TMDCA.

\section{Method}
\label{sec:formalism}
We consider the Anderson model of non-interacting electrons subjected to a random potential. The Hamiltonian (for 
spinless fermions) is given by 
\begin{equation} \label{eqn:model}
H=-\sum_{<ij>}t_{ij}(c_{i}^{\dagger}c_{j}+h.c.)+\sum_{i}(V_i-\mu) n_{i}.
\end{equation}
The first term is the kinetic energy operator due to hopping of electrons on a lattice. 
The operators $c_{i}^\dagger$($c_{i}$) create (annihilate) a quasiparticle 
on site $i$, and $t_{ij}$ is the 
hopping matrix element between nearest neighbors $\left<i,j\right>$. 
The second term is the disordered part parameterized by a local potential $V_i$, which is a random quantity distributed
according to some specified probability distribution $P(V_i)$, $n_{i} = c_{i}^\dagger c_{i}$ is the number operator, 
and $\mu$ is the chemical potential. We set $4t = 1$ as the energy unit. In our analysis, we use different disorder distributions. 
In particular, we consider the box (Bo), Gaussian (Ga), Lorentzian (Lo), and binary (Bi) distributions, 
respectively, with the corresponding distribution functions $P(V_i)$
\begin{subequations} 
\begin{align} 
  P_{Bo}(V_i)&=\frac{1}{2W_{Bo}}\Theta(W_{Bo}-|V_i|), \label{eqn:distribution_Box} \\
  P_{Ga}(V_i)&=\sqrt{\frac{3}{2\pi W_{Ga}^2}}e^{-3V_i^2/(2W_{Ga}^2)}, \label{eqn:distribution_Gaus} \\
  P_{Lo}(V_i) &=\frac{W_{Lo}}{\pi (V_i^2+W_{Lo}^2)}, \label{eqn:distribution_Lo}\\ 
  P_{Bi}(V_i) &=c_a\delta(V_i-W_A)+c_b\delta(V_i-W_B), \label{eqn:distribution_Bi}
\end{align} 
\end{subequations}
where $\Theta(x)$ is the step function, $c_a$ is the concentration of the host A atom, $c_b =1-c_a$ 
is the concentration of the impurity B atom, and the strength of the 
disorder in units of $4t$ is parameterized by $W$ (W$_A$ and W$_B$ for the  binary alloy model). We 
have scaled W$_{Ga}$ such that the second moments of the P$_{Bo}$ and P$_{Ga}$ agree with 
each other (i.e., we set the variance of the Gaussian distribution equal to that of the box 
distribution: $\sigma^2=W^2/3$) in the event that W$_{Bo}$=W$_{Ga}$ to enable comparison. 
Since the Lorentzian distribution lacks a second moment, the disorder values cannot be directly 
compared with that of either the box or Gaussian distributions.  We introduce a
shorthand notation for disorder averaging: $\langle...\rangle=\int dV_i P(V_i) (...)$. 

To solve the Hamiltonian (\ref{eqn:model}), different methods will be used including the 
DCA,~\cite{PhysRevB.63.125102} the cluster typical medium theory,~\cite{0953-8984-26-27-274209} and 
the recently developed typical medium dynamical cluster approximation (TMDCA).~\cite{PhysRevB.89.081107}  
We will compare these results to those obtained from numerical methods like the 
kernel polynomial method (KPM) \cite{Schubert,MacKinnonKramer1983,Kramer2010,Markos-review-2006} 
and the transfer matrix method (TMM).~\cite{PhysRevLett.47.1546,Kramer,Kramer1996,MacKinnonKramer1983,Kramer2010,Markos-review-2006}

The TMDCA utilizes the self-consistent framework of the standard dynamical cluster 
approximation~\cite{PhysRevB.63.125102} with the important usage of an environment defined by a 
typical nonlocal hybridization function. In particular, the TMDCA maps the given disordered lattice 
system onto a finite cluster which is embedded in an effective self-consistent typical medium. Note that 
unlike the usual DCA scheme, where the effective medium is constructed via algebraic averaging over 
disorder configurations, the TMDCA scheme uses geometric averaging. By mapping a $d$-dimensional lattice 
to a finite small cluster containing $N_c$ $=$ $L^d_c$ sites, where $L_c$ is the linear 
dimension of the cluster, we dramatically reduce the computation effort.\cite{RevModPhys.77.1027} 
Unlike the single-site methods commonly used to study disordered systems, such as the coherent potential 
approximation (CPA) \cite{RevModPhys.46.465,PhysRev.156.809} or the local TMT, \cite{Vlad2003} the TMDCA 
ensures that nonlocal spatial fluctuations, neglected in single-site approaches, are systematically 
incorporated as the cluster size $N_c$ increases. Short length scale correlations are
treated exactly inside the cluster, while the long length 
scale correlations are treated within the typical medium. 

\textit{Algorithm}: The details of the TMDCA formalism are 
described below. The nonlocal ($\K$-dependent) disorder average cluster density of states  is given as
\begin{equation} \label{eqn:3}
\rho_{avg}^c(\K,\omega)= \langle \rho^c(\K,\omega,V)\rangle = -\frac{1}{\pi} \langle \Im  G^c (\K,\K,\omega,V) \rangle,
\end{equation}
where the superscript `$c$' denotes cluster and $\left\langle ... \right\rangle$ is the disorder average.
For a single-site $N{}_c = 1$, we recover the CPA. The $\K$-dependent cluster 
Green function is obtained from the site dependent Green function $G_c(i,j,\omega)$ via the Fourier transform:
\begin{equation}
G^c (\K,\K,\omega) = \frac{1}{N_c} \sum_{i,j} e^{i\K\cdot (\R_i-\R_j)} G_c(i,j,\omega).
\end{equation}
Within the TMDCA, for each cluster configuration, we first obtain 
$\rho^{c}(\K,\omega)=-\Im G^c (\K,\K,\omega)/\pi$. It can be shown via 
the Lehmann representation,\cite{Gross1991,fetter1971quantum} 
that $\rho^{c}(\K,\omega)\geq 0$ for each $\K$, $\omega$, and disorder configuration. 

As mentioned above, in the TMDCA, the local part of the cluster-momentum-resolved typical density of states is 
separated and treated with geometrical averaging over the disorder 
configurations, to avoid self-averaging as the cluster size increases. The obtained cluster typical spectra are given by

\begin{widetext}
\begin{equation}\label{Eq:rho_typ_definition}
\rho_{typ}^c(\K,\omega)
=
\longoverbrace{\exp\left(\frac{1}{N_c} \sum_{i=1}^{N_c} \left\langle \ln \rho_{i}^c (\omega,V)  \right\rangle\right)}{local TDOS} \times 
\longunderbrace{\left\langle \frac{\rho^c(\K,\omega,V)}{\frac{1}{N_c} \sum_{i} \rho_{i}^c (\omega,V)} \right\rangle }{nonlocal}.
\end{equation}
\end{widetext}

\begin{figure}[b!]
\begin{center}
 \includegraphics[trim = 0mm 0mm 0mm 0mm,width=1.0\columnwidth,keepaspectratio,clip=true]{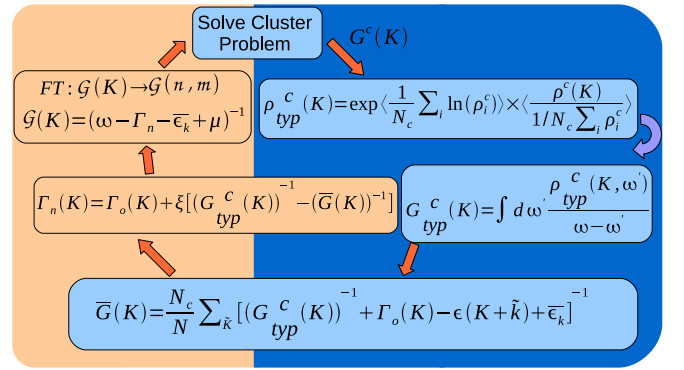}
\caption{(Color online) The self-consistent loop of the TMDCA.}
\label{algorithm-self_consistency}
\end{center}
\end{figure}

%
From Eq.~\ref{Eq:rho_typ_definition}, the disorder averaged typical cluster Green function is obtained 
using the Hilbert transform 
\begin{equation}\label{eqn:HilbertT}
G_{typ}^c(\K,\omega)=\int d \omega' \displaystyle \frac{\rho_{typ}^c(\K,\omega')}{\omega - \omega'}.
\end{equation}
A schematic TMDCA self-consistency is shown in Fig.~\ref{algorithm-self_consistency}. 

The TMDCA iterative procedure is outlined below.
\begin{enumerate}
 \item 
The TMDCA iterative procedure begins by proposing an initial guess for the hybridization function $\Gamma_o (\K,\omega)$, 
where the subscript `$o$' denotes old. The choice of the starting guess for the hybridization function may 
be based on a priori knowledge, i.e., having information about the self-energy $\Sigma(\K,\omega)$ 
and cluster Green function $G^c(\K,\omega)$, $\Gamma_o (\K,\omega)$ can be calculated as
\begin{multline} \label{eqn:oldhybrid}
\Gamma_o (\K,\omega) = \omega - \overline{\epsilon}(\K) + \mu - \Sigma (\K,\omega) - 1/G^c(\K,\omega),
\end{multline}
where $\overline{\epsilon}(\K)$ $=$ $\displaystyle N_c/N\sum_{\tilde{k}}{\epsilon(\K+\tilde{k})}$ is the 
coarse-grained bare dispersion with $\tilde{k}$ summed over $N/N_c $ momenta 
inside the cell centered at the cluster momentum $\K$.~\cite{PhysRevB.61.12739} However, if nothing is 
known a priori, $\Gamma_o (\K,\omega)$  set to a small imaginary number may serve as the starting point.

\item
After setting up the cluster problem, we calculate the cluster-excluded Green function 
${\cal G}(\K,\omega)$ as

\begin{equation} \label{eqn:clusterexG}
{\cal G}(\K,\omega)
=
(\omega - \Gamma_o (\K,\omega) - \overline{\epsilon}(\K)+ \mu)^{-1}.
\end{equation}

Since the cluster problem is solved in real space, we then Fourier transform ${\cal G}(\K,\omega$):
${\cal G}_{n,m} = \sum_{\K} {\cal G}(\K)\exp(i \K\cdot(r_n-r_m))$.

\item
Next, to solve the cluster problem, we stochastically
generate random configurations of the disorder potential $V$, and calculate the 
corresponding cluster Green function as
\begin{equation}
G^c(V) = ({\cal G}^{-1} - V)^{-1}.
\label{eq:GcofV}
\end{equation}
This is Fourier transformed to $G^c(\K,\K,\omega)$ to obtain the cluster density of states 
$\rho^c(\K,\omega)=-\frac {1}{\pi} \Im G^c(\K,\K,\omega)$.  The typical cluster spectra 
is then calculated via geometric averaging using Eq.~\ref{Eq:rho_typ_definition}. Then,
we calculate the disorder averaged, typical cluster Green function $G_{typ}^c(\K,\omega)$ 
via the Hilbert transform using Eq.~\ref{eqn:HilbertT}.  We note the advantage of the stochastic 
sampling of the disorder configurations. Here, each of the disorder configurations is 
statistically independent of the others. Thus, for example, for the binary disorder distribution,
instead of enumerating all configurations, which scales as 2$^{Nc}$, we do a stochastic sampling 
of the disorder configurations. This greatly reduces the computational cost, at the expense of 
a small sampling error, enabling us to study larger clusters.  We also enforce all of the cluster
translational and point group symmetries, effectively generating more configurations.  With this,
the number of disorder realizations needed to obtain a converged solution falls with increasing 
cluster size. For a typical 64 site cluster, with box disorder, about 500 disorder realizations 
are enough to produce high-quality data.  The code scales like $A(N_c) N_c^3$ due to the matrix 
inversion in Eq.~\ref{eq:GcofV}, but the prefactor $A(N_c)$, also depends on $N_c$ since fewer 
self-consistency iterations and disorder configurations are needed for larger clusters. Hence, 
$A(N_c)$ falls with increasing cluster size.

\item
After solving the cluster problem, we use the typical cluster Green function, $G_{typ}^c(\K,\omega)$, 
to calculate the coarse-grained cluster Green function of the lattice $\overline{G} (\K,\omega)$ as
\begin{equation}
\overline{G} (\K,\omega)  
= \int \displaystyle \frac{N^c_0(\K,\epsilon) d\epsilon}{(G^c_{typ} (\K,\omega))^{-1} + \Gamma (\K,\omega) - 
 \epsilon + \overline{\epsilon}(\K) },
\end{equation}
where $N^c_0(\K,\epsilon)$ is the bare partial density of states.

\item
We then close our self-consistency loop by updating the new hybridization function using linear mixing

\begin{equation} \label{eqn:newhybrid}
\Gamma_n (\K,\omega)
=
\Gamma_o (\K,\omega) + \xi [(G^c_{typ} (\K,\omega))^{-1} - (\overline{G} (\K,\omega))^{-1}]
\end{equation}
where the subscripts `$n$' and `$o$' denote new and old, respectively. The mixing parameter 
$\xi > 0 $ controls the ratio of the new and old $\Gamma(\K,\omega)$ entering the next iteration. 
For very small $\xi$, convergence may be slowed down unnecessarily, while for very 
large $\xi$, oscillations about the self-consistent solution may occur. Instead of 
linear mixing, the convergence of the computations can be improved by using the 
Broyden method.~\cite{PhysRevB.38.12807}

\item
We repeat the above procedure until the hybridization function converges to the desired 
accuracy, $\Gamma_o(\K,\omega)=\Gamma_n(\K,\omega)$. When this happens, the Green functions 
are also converged, $\overline{G}(\K,\omega)= G^c_{typ}(\K,\omega)$ within the computational error. 
\end{enumerate}

We note that our formalism  preserves causality just as the DCA,~\cite{PhysRevB.63.125102} 
since all the Green functions are causal, both the average density of states (ADOS) and the TDOS calculated from them are 
positive definite. Also, we observe that as $N_c$ increases, our method systematically 
interpolates between the local TMT and the exact result.

For reproducibility, we specify in Table~\ref{tab:clus3d} the cluster geometries and other 
important parameters of the clusters used in our computations. 
The parameters of Table~\ref{tab:clus3d} include the lattice vectors $(\vec{a}_1 , \vec{a}_2 , \vec{a}_3)$, and the 
cubicity (C).~\cite{PhysRevB.72.060411} 
The cubicity is given as  
$C=\max(c_1,c_1^{-1})\times \max(c_2,c_2^{-1})$, where $c_1 = 3^{1/2}l/d$ and $c_2 = 2^{1/2}l/f$ are cluster 
parameters defined by the geometric mean of the lengths of the four body diagonals of the cluster, 
$d=\left(d_1d_2d_3d_4\right)^{1/4}$, the six-face diagonals, $f = (f_1f_2f_3f_4f_5f_6)^{1/6}$, and 
the edges, $l = (l_1l_2l_3)^{1/3}$.~\cite{PhysRevB.72.060411} $C=1$ is for a perfect cube, and $C>1$ otherwise. Following 
this criterion, clusters $N_c=1$, 64, 125, and 216 are perfect cubes. 
\begin{table}[htb] 
\centering
\caption{Three-dimensional (3D) cluster geometries utilized in our calculations.  The $a_{i}$ denote the cluster 
lattice vectors and $C$ is the cubicity.} 
\begin{tabular}{cccccccc}
\hline\hline
$N_c$ & $\vec{a}_1$ & $\vec{a}_2$ & $\vec{a}_3$ & C \\ 
\hline 
1&(1, 0, 0)& (0, 1, 0)& (0, 0, 1)&1.000 \\
38&(1, 2, 3)& (3, -1, -2)& (2, -2, 2)&1.087 \\
64&(4, 0, 0)& (0, 4, 0)& (0, 0, 4)&1.000 \\
125&(5, 0, 0)& (0, 5, 0)& (0, 0, 5)&1.000 \\
216&(6, 0, 0)& (0, 6, 0)& (0, 0, 6)&1.000 \\
\hline \hline
  \end{tabular}
  \label{tab:clus3d}
\end{table}

\section{Results and Discussion}
\label{sec:results}
Before presenting our main results in detail, we will first review the characteristics of the DCA.  
Despite its advantage over the CPA, it shares the same behavior with the CPA in that it is unable to detect the localization 
transition in a disordered electron system.\cite{Jarrell01} We will also elaborate on the details of how self-averaging 
is avoided in the TMDCA as the cluster size is increased. This becomes imperative since in the cluster, 
self-averaging will ultimately destroy our ability to detect the Anderson localization transition. Both 
the inability of the DCA to capture the Anderson localization transition and how self-averaging is avoided 
in the TMDCA will be demonstrated using the box disorder distribution.

\subsubsection{Absence of Localization in the DCA}
\label{sec:absence_DCA}
\begin{figure}[b!]
 \includegraphics[trim = 0mm 0mm 0mm 0mm,width=1\columnwidth,clip=true]{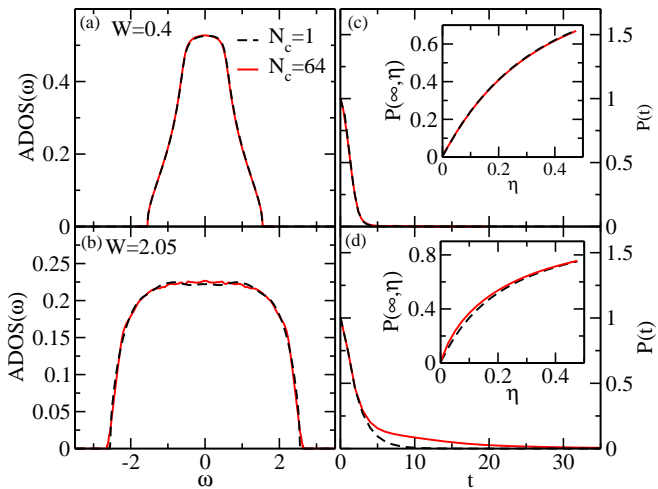}
\caption{(Color online). Left panels: The average density of states calculated using the DCA for small and 
large disorder strengths\cite{Jarrell01}. Right panels: The probability of an electron remaining on a site after 
time $t$, $P(t) = \left\langle | G(l,l,t) |^2 \right\rangle$. The insets in the plots on the right are 
the probability of an electron remaining on a site for all time 
$P(t\rightarrow\infty,\eta) = \lim\limits_{t \to \infty} \left\langle  | G(l,l,t) |^2  \right\rangle =$ 
$=\lim\limits_{\eta \to 0} \frac{\eta}{\pi} 
\int_{-\infty}^{\infty} d\epsilon \left\langle | G(l,l,\epsilon+i\eta) |^2 \right\rangle$
for the $N_c=1$ (black) and 
64 (red), respectively. The ADOS is not critical at the Anderson transition. The non-critical behavior in the DCA can 
also be inferred from P$(t)$, which decays rather fast instead of remaining constant, and in that $P(t,\eta)$ 
extrapolates to zero instead of a finite value when many of the states are localized.
The DCA only shows a precursor to the Anderson localization as manifested in P$(t)$ for large disorder 
strength for the finite cluster sizes.} 
\label{Fig3:pt-dca}
\end{figure}

The dynamical cluster approximation (DCA), unlike the coherent potential approximation (CPA) incorporates nonlocal 
spatial correlations systematically as the size of the cluster is increased. While spatial correlations are an 
important ingredient in the localization transition in disordered electron systems, the DCA effective medium is 
characterized by arithmetic averaging over the disorder configurations. As explained above and will be demonstrated 
below, even a typically defined medium without a proper treatment of the typical density of states (the local part 
of the typical density of states needs to be separated and treated explicitly using geometrical averaging), reduces 
to the DCA for large clusters. In Fig.~\ref{Fig3:pt-dca} (left panel), we show the ADOS at small and large disorder 
strengths for the single-site ($N_c=1$) and finite cluster ($N_c=64$) calculations. We also show in 
Fig.~\ref{Fig3:pt-dca} (right panel) the probability $P(t) = \left\langle | G(l,l,t) |^2 \right\rangle$  
of an electron remaining at a site $l$ at long time $t$ calculated within the single-site and finite cluster 
DCA~\cite{Jarrell01} scheme while the insets depict the probability of an electron remaining on a site $l$ 
for all time: 
$P(t\rightarrow\infty,\eta) = \lim\limits_{t \to \infty} \left\langle  | G(l,l,t) |^2  \right\rangle =$ 
$=\lim\limits_{\eta \to 0} \frac{\eta}{\pi} 
\int_{-\infty}^{\infty} d\epsilon \left\langle | G(l,l,\epsilon+i\eta) |^2 \right\rangle$. 

$P(\infty,\eta)$ is expected ~\cite{Jarrell01} to be nonzero for any fraction of the localized states in the spectrum 
of the eigenstates of the disordered system.~\cite{PhysRevB.5.4802,a_gonis_92,Hartmann1989,PhysRevB.63.125102} As is evident 
from the inset of the right panel of Fig.~\ref{Fig3:pt-dca}, the plot of $P(\infty,\eta)$ versus $\eta$, 
$P(\infty,\eta)$ extrapolates to zero even very close to the critical disorder strength for both $N_c=1$ and 64, 
respectively.  

Also observe from Fig.~\ref{Fig3:pt-dca} (left panel), the ADOS is not critical at the Anderson transition. 
This is manifested in the $P(t)$ plot since for a localized state, $P(t)$ is expected to be 
finite. For $N_c=1$, $P(t)$ falls quickly with time regardless of the disorder strength. However, for $N_c=64$, 
the electrons remain localized for longer times as the disorder strength is increased. 
This can be understood by noting that each site on the cluster is coupled to a non-interacting 
translationally invariant host into which electrons can escape. Hence, if a finite density of states 
exists at some energy, the corresponding states are guaranteed to be extended unless the hybridization rate 
between the host and the cluster vanishes. Indeed, this is the case in the DCA. The imaginary part of the 
integrated hybridization (escape) rate ($-\Im \int \Gamma(\K,\omega) d\K d\omega$) between the cluster and the host 
as a function of disorder strength ($W$) remains constant regardless of the strength of the 
disorder (cf.\ insets of Figs.~\ref{Fig2:tdos_vs_V}, \ref{Fig:TDOS_Disorder-Gaussian} and ~\ref{Fig:TDOS_Disorder-LD},
for different disorder distributions). 
It does not go to zero as needed for a localization transition in the typical medium context.
Thus, the DCA is only able to capture the precursor to the Anderson localization as the cluster size, but not the disorder
strength increases. 

\subsubsection{Avoiding Self-averaging}
\label{sec:self-averaging}
The averaging procedure used to calculate the typical spectra is not unique. As noted above, 
our initial attempt to formulate a cluster version of the TMT reproduced the expected behavior  
in one and two dimensions as $N_c$ is increased.~\cite{0953-8984-26-27-274209} However, in three dimensions, 
applying the algorithm directly will lead to an effective self-averaging for large clusters. 
This is due to the fact that close to criticality, there exist distinct localized and extended states 
above and below the localization edge given by the TDOS with an energy scale difference that can span an 
order of magnitude.  These energy scales need to be treated differently. This can be seen by investigating the spectra 
where the local part of the TDOS is not explicitly separated and treated with a geometric averaging 
over disorder realizations:
\begin{equation} \label{rho_typ}
\rho_{typ}^c(\K,\omega)
=
\exp \left\langle \ln \rho^c (\K,\omega,V_i) \right\rangle. 
\end{equation}
In forming the Fourier transform
\begin{eqnarray} \label{rho_FT}
\rho^c (\K,\omega,V_i) = -\frac{1}{\pi} \Im \left[\frac{1}{N_c}\sum_{X,X'} e^{i\K\cdot(X-X')}G^c (X,X',\omega,V_i)\right] \nonumber \\ 
\end{eqnarray}
we average over the cluster coordinates $X$ and $X'$, including the local part, $X=X'$.  So, the local 
DOS is first averaged over the cluster sites and then Fourier transformed making the local part of $\rho^c(\K,\omega)$.  
Hence, for large clusters, this reduces to linear averaging of the local part instead of geometrical 
averaging. As a consequence, the host Green function constructed from $\rho_{typ}^c(\K,\omega)$ of Eq.~\ref{rho_typ}
is unaware of the TDOS and thus, it is unable to distinguish between the energies above and below the localization edge. 
To avoid such self-averaging in the TDOS, we proposed the Typical Medium DCA 
(TMDCA) method.~\cite{PhysRevB.89.081107} Here, the cluster-momentum-resolved typical density of states (TDOS) for each $\K$ is split 
into local and nonlocal parts. The local part is treated with geometrical averaging over disorder configurations, 
while the nonlocal part is treated with an algebraic or geometric averaging over the disorder 
configuration.

To do this, we have utilized two schemes. The first scheme is what we call \textit{linear-log} procedure, which 
is used in this study. Here, we treat the local part with a geometrical averaging while the 
nonlocal part is approximated algebraically using linear averaging as
\begin{multline} \label{Eq:log-linear}
\rho_{typ}^c(\K,\omega)
=
\exp\left(\frac{1}{N_c} \sum_{i=1}^{N_c} \left\langle \ln \rho_{i}^c (\omega,V_i)  \right\rangle\right) \times \\ 
\left\langle \frac{\rho^c(\K,\omega,V_i)}{\frac{1}{N_c} \sum_{i} \rho_{i}^c (\omega,V_i)} \right\rangle.
\end{multline}
The second scheme is what we call the \textit{log-log} procedure, which again involves the treatment of 
the local part with geometrical averaging; however, the nonlocal part is treated with a log averaging as 
\begin{multline} \label{Eq:log-log}
\rho_{typ}^c(\K,\omega)
=
\exp\left(\frac{1}{N_c} \sum_{i=1}^{N_c} \left\langle \ln \rho_{i}^c (\omega,V_i)  \right\rangle \right) \times \\ 
\exp \left( \left\langle \ln \frac{\rho^c(\K,\omega,V_i)}{\frac{1}{N_c} \sum_{i} \rho_{i}^c (\omega,V_i)} \right\rangle \right).
\end{multline}

\begin{figure}[th!]
 \includegraphics[trim = 0mm 0mm 0mm 0mm,width=1\columnwidth,clip=true]{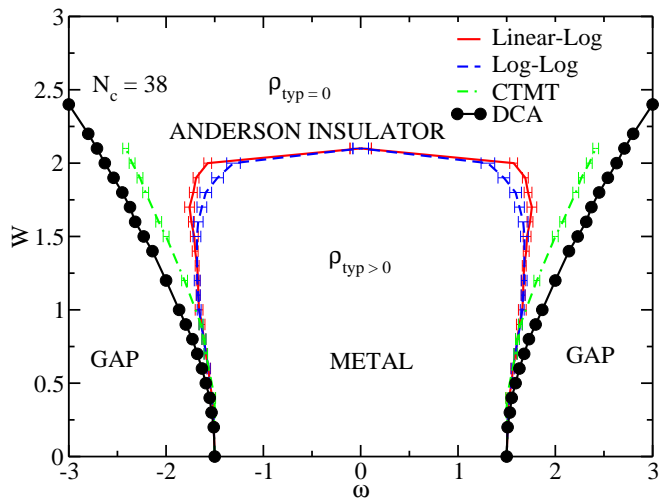}
\caption{(Color online). A comparison of the phase diagrams of the Anderson localization transition in 3D 
obtained  from different cluster approximations with $N_c=38$ using the CTMT and the TMDCA (linear-log 
and log-log) schemes. Observe that in the CTMT, as a consequence of self-averaging, the higher disorder 
behaviors that are captured in our TMDCA are totally missed and the critical disorder strength is also 
severely over-estimated.}
\label{Fig:Phase-Diagram_Nc38-compare.eps}
\end{figure}

It is imperative to note that while there are different behaviors of the two methods around the re-entrance 
region (cf. Fig.~\ref{Fig:Phase-Diagram_Nc38-compare.eps}), both procedures systematically converge to the same critical disorder strength 
e.g., $W_c^{N_c\geq12}\approx 2.1\pm 0.01$ for the box disorder distribution. 
However, the former (\textit{linear-log} procedure) is generally more robust 
than the \textit{log-log} method. The latter is characterized by slower convergence around the re-entrance region, 
requiring far larger cluster sizes before the convergence of the re-entrance region is achieved in comparison 
to the e.g., the transfer matrix method (TMM) results. It may also not be adequate to study localization phenomena in 
realistic material applications, since it is not obviously clear how a geometrical averaging  of 
the off-diagonal components of the spectral density, which are not positive definite, will be done. 
The comparison of the phase diagram obtained using the two procedures: \textit{log-log} and \textit{linear-log} formalisms 
is shown in Fig.~\ref{Fig:Phase-Diagram_Nc38-compare.eps}. As is evident from 
the figure, the two new schemes converge to the same 
critical disorder strength but behave differently around the re-entrance region. While our original formulation 
(cluster typical medium theory (CTMT)) will eventually 
converge to a disorder strength far greater than $W_c$, a further remark is that the re-entrance trajectory of the 
mobility edge is totally missed as a consequence of self-averaging in the cluster. 

\medskip{}

We note that in both the \textit{linear-log} and the \textit{log-log} procedures, at small $N_c$, about 
100 self-consistent iterations are required 
to achieve convergence, while, for relatively large $N_c$, far fewer iterations are required. The convergence 
criterion in both limits is achieved when the TDOS\,($\omega = 0$) does not fluctuate anymore with iteration number 
within the error bars. 

Finally, we note that many other definitions of the typical medium which avoid self-averaging are possible, 
including the use of only the local part of Eq.~\ref{Eq:log-linear}, i.e.,
\begin{equation}
\rho_{typ}^c(K,\omega)
=
\exp\left(\frac{1}{N_c} \sum_{i=1}^{N_c} \left\langle \ln \rho_{i}^c (\omega,V_i)  \right\rangle\right) \,.
\end{equation}
However, this method was rejected since it does not meet most of the criteria discussed in the 
introduction. In this case, this formalism does not recover the DCA in the weak disorder limit.

\subsubsection{The Pole Procedure}
\label{sec:pole_procedure}
Close to criticality, the hybridization rate of the states at the top and bottom (($0,0,0$) and ($\pi,\pi,\pi$)) 
of the bands tends to zero leading to the development of poles 
in the cluster excluded Green function. Here, we present in detail how to deal with such poles that 
emerge on the real frequency axis as the critical disorder strength is approached.

When $\Im \Gamma_{typ} (\K,\omega)$ becomes very small (i.e., $\Im \Gamma_{typ} (\K,\omega)\rightarrow 0$), the imaginary part of the 
cluster-excluded Green function, 
${\cal G}(\K,\omega)$, becomes a series of delta functions. To see this, we note that
\begin{eqnarray} \label{eqn:cluster_pole}
{\cal G}(\K,\omega)
&=&  (\omega - \Gamma_{typ} (\K,\omega) - \overline{\epsilon}(\K)+\mu)^{-1} \\
& =& \specialp (\omega-\omega^\prime)^{-1} -i\pi \delta(\omega -\omega^\prime ) \nonumber,
\end{eqnarray} 
where $\omega^\prime =  \overline{\epsilon}(\K) -\mu + \Re \Gamma_{typ} (\K,\omega)$ and ``\specialp'' denotes 
the principle value.  Evidently from Eq.~\ref{eqn:cluster_pole}, the poles cannot be represented 
in the conventional way as a list of frequencies on the computer with a finite frequency 
resolution $d\omega$. Such difficulty can be avoided by replacing ${\cal G}(\K,\omega)$ for each of the $\K-$cells 
where $\Im \Gamma_{typ} (\K,\omega)$ is vanishing with 
\begin{equation}
{\cal G}(\K,\omega) = \left\{ \begin{array}{r@{\quad:\quad}l}
-i\pi/{d\omega} &  \omega=\omega^\prime\\
\frac{1}{\omega-\omega^\prime} & \omega \neq \omega^\prime.
\end{array} \right.
\end{equation}
We refer to this formulation as the explicit ``\textit{pole-procedure}''. 
With this procedure, the singularity in ${\cal G}(\K,\omega)$ can be properly captured. 
An added difficulty is that for a given $N_c$, as the $W_c$ is approached, $\Im \Gamma_{typ} (\K,\omega)$ 
for individual cells goes to zero at different rates. Hence, we have to determine which of these cells need to 
be treated with the explicit ``pole-procedure''. We choose the criterion that for any cell, if 
$(-1/\pi)\times\Im\Gamma_{typ} (\K,\omega^\prime)$ $<$ $a\times d\omega^\prime$, then, we apply the 
pole procedure to such cells. Here, $a \agt 1$ is a parameter that measures the minimum number of 
pixels needed to represent a pole approaching the real frequency axis. Our numerical experience shows 
that such a criterion works nicely while spurious results are obtained otherwise.

\subsection{Box Disorder Distribution}
\label{sec:boxdistribution}
\begin{figure}[b!]
 \includegraphics[trim = 0mm 0mm 0mm 0mm,width=1\columnwidth,clip=true]{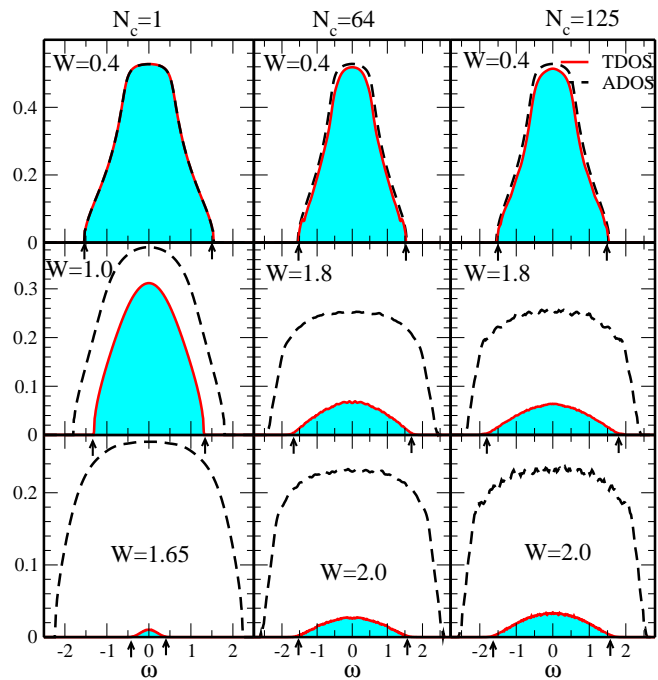}
\caption{(Color online). The evolution of the ADOS and TDOS at various disorder strengths $W$ 
for the single-site TMT and the TMDCA with cluster size $N_c=64$ and 125. At low disorder, 
where all the states are metallic, the shape of TDOS is the same as that of the ADOS. 
As $W$ increases, in the case of single-site TMT, the TDOS gets suppressed and 
the mobility edge moves towards $\omega=0$ monotonically. In the TMDCA, 
the TDOS is also suppressed, but the mobility edge 
first moves to higher energy,and  only with a further increase of $W > 1.8$, 
it starts moving towards the band center, 
 indicating that the TMDCA can successfully capture the re-entrance behavior.
 Arrows indicate the position of the mobility edge, which separates 
the extended electronic states from the localized ones and the colored region indicates the TDOS. } 
\label{Fig1:tdos_ados}
\end{figure}
To demonstrate that the typical and not the average DOS can serve as a proper order parameter 
for the Anderson localization transition, we start the discussion of our results by comparing the 
algebraically averaged DOS (ADOS) calculated using DCA and the TDOS obtained from a single-site TMT ($N_c=1$) and finite 
clusters TMDCA ($N_c=64$ and 125) at various disorder strengths $W$ for the box 
disorder distribution (Eq.~\ref{eqn:distribution_Box}). As shown in
Fig.~\ref{Fig1:tdos_ados}, the ADOS remains finite while the TDOS for both TMT and TMDCA
continually gets suppressed as the critical disorder strength is approached. 
Moreover, one observes a crucial difference between the single-site TMT ($N_c=1$) and TMDCA 
finite clusters of $N_c=64$ and 125. 
In the former, the mobility edge (for extended states TDOS is finite)
defined by the boundary of the TDOS (indicated by arrows) always gets narrower with increasing disorder strength $W$,
while in the latter, as a function of disorder strength, the mobility edge 
first expands and then decreases, hence giving rise to the re-entrance behavior, which 
is completely missing in the single-site TMT. Observe also the quick convergence with the clusters size 
at $\omega=0$ for the finite clusters $N_c=64$ and 125. The implications of this will be discussed 
further below and explored with respect to what happens to the trajectories of the mobility edge as the 
size of the cluster increases.

\begin{figure}[t!]
\includegraphics[trim = 0mm 0mm 0mm 0mm,width=1\columnwidth,clip=true]{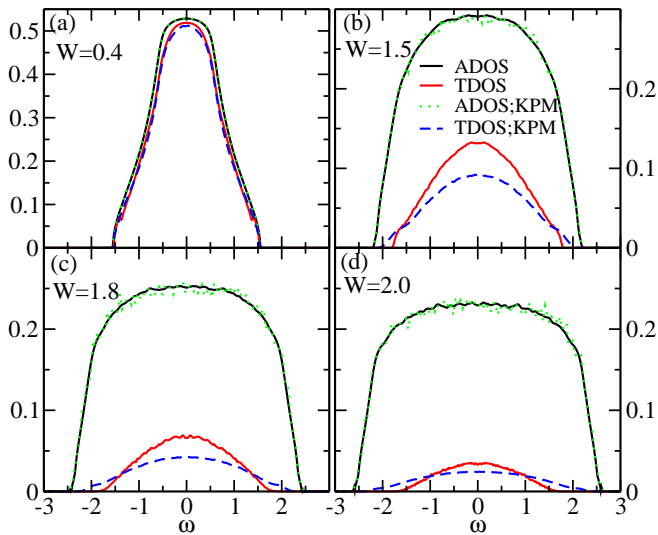}
\caption{(Color online). Comparison of the average (typical) density of states calculated with the DCA 
(TMDCA) and the kernel polynomial method (KPM) for the box disorder distribution at various strengths 
for the cluster size $N_c$=64. The kernel polynomial method used 4096 moments on a $48^3$ cubic 
lattice, and 1000 independent realizations generated with 32 sites randomly sampled from each realization. }
\label{Fig:TMDCA_KPM_compare}
\end{figure}

To benchmark our TMDCA formalism with another numerical technique, we utilize the kernel 
polynomial method.~\cite{KPM_review_2006, Schubert} We show in Fig.~\ref{Fig:TMDCA_KPM_compare} a 
plot of the TDOS (ADOS) ($N_c=64$) 
calculated using the TMDCA (DCA) as compared to the TDOS (ADOS) obtained using the kernel polynomial method (KPM). 
As it is evident from the plots, even though there is a qualitative agreement between the two methods, there 
are subtle deviations especially in the TDOS. This deviation can be attributed to finite lattice effects and 
the effective broadening due to the finite order expansion used in the KPM. Overall, the agreement is a manifestation 
of the ability of our TMDCA formalism~\cite{PhysRevB.89.081107} to accurately characterize the Anderson localization 
transition in systems with a uniform disorder distribution even with relatively small system sizes, as compared to the 
large lattice systems that need to be simulated in the kernel polynomial method for accurate results to be obtained.

\begin{figure}[t!]
\includegraphics[trim = 0mm 0mm 0mm 0mm,width=1\columnwidth,clip=true]{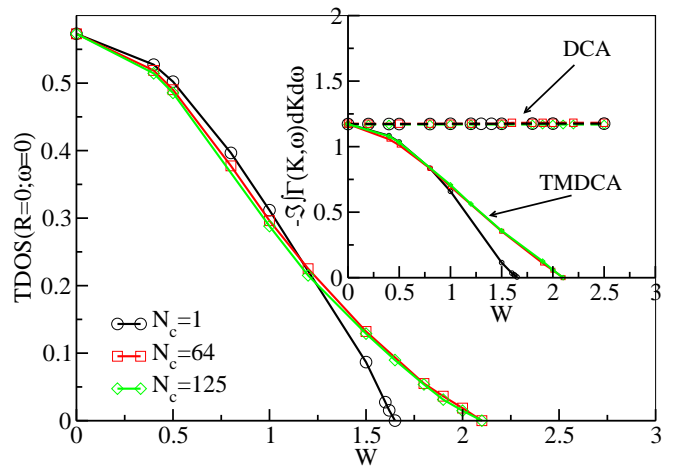}
\caption{(Color online). The TDOS$(\omega=0)$ vs. disorder strength $W$ at different cluster sizes $N_c=1,64,$ 
and 125 for the uniform (box) disorder distribution. The TDOS$(R=0,\omega=0)$ vanishes at the critical disorder 
strength $W_c$.  At $N_c=1$,  $W_c^{N_c=1} \approx 1.65$. 
As the cluster size increases, $W_c$ systematically increases with 
$W_c^{N_c\gg12}\approx 2.10\pm 0.10$, showing a quick convergence with the cluster size. The inset shows 
the integrated hybridization function ($-\Im \int \Gamma(\K,\omega) d\K d\omega$) as a function of disorder 
strength $W$. Observe that $-\Im \int \Gamma(\K,\omega) d\K d\omega$ vanishes at the same disorder strength 
as the TDOS. The dashed lines are the $-\Im \int \Gamma(\K,\omega) d\K d\omega$ from the DCA. Observe that it is 
a constant regardless of the disorder strength and cluster size. This shows that the DCA, even though it incorporates 
spatial correlations, does not describe the Anderson localization transition. Moreover, near the critical 
region the TDOS$(R=0,\omega=0)$ data can be fitted to a power-law, with TDOS$(R=0,\omega=0)=a_0|W-W_c^{fit}|^\beta$. 
The obtained critical exponent for large  enough clusters $\beta \approx 1.62 \pm 0.10$ is in good agreement
with exact results. Note that the $-\Im \int \Gamma(\K,\omega) d\K d\omega$ data for $N_c=64$ and 125 has been normalized 
with that of $N_c=1$.}
\label{Fig2:tdos_vs_V}
\end{figure}

Next, we consider the evolution of the critical disorder strength $W_c$ with the cluster size.
Figure~\ref{Fig2:tdos_vs_V} shows the local TDOS($\omega=0$) at the band center as a function of disorder strength $W$ 
for several cluster sizes: $N_c=1, 64,$ and 125. The critical disorder strength $W_c$ is defined by the 
vanishing of the TDOS($\omega=0$). The inset is the imaginary part of the integrated hybridization function
which shares the same property as the TDOS since both vanish at the same disorder strength, while the DCA
result remains finite, independent of the disorder strength, indicating no tendency towards localization with 
increasing disorder. Our results show that as $N_c$ increases, the critical disorder strength $W_c$ in the 
TMDCA systematically increases until it converges to the exact value $W_c\approx2.10$~\cite{Slevin99,Bulka85,Fehske,PhysRevB.76.045105,
PhysRevB.63.045108,PhysRevB.84.134209,PhysRevLett.105.046403,PhysRevLett.47.1546} at cluster size 
$N_c\ge12$. The cluster $N_c=12$ is the first cluster with a complete nearest-neighbor shell based 
on Betts cluster classification.~\cite{betts-3d} From this cluster onward, $W_c$ converges to $\approx$ 
2.10, but the trajectory of the mobility edge in the re-entrance regime continues 
to change until it also converges to the exact results at larger $N_c$. This effect is due to the 
systematic incorporation of coherent backscattering as the cluster size increases and 
will be elaborated in more detail later. 

To extract the order parameter critical exponent ($\beta$), 
we fit our data in Fig.~\ref{Fig2:tdos_vs_V} for the largest system size considered here ($N_c=125$) 
using the power law: TDOS$(\omega=0)=a_0|W-W_c^{fit}|^\beta$. Following the procedure as explained in 
Sec.~\ref{criticalparameters}, we obtain $\beta\sim$ $1.62\pm0.10$ with a corresponding critical 
disorder strength from the fit $W_c^{fit}\sim2.23\pm0.10$. The fit overestimates the critical disorder
strength, as compared to the computed one, due to the difficulty in determining 
the scaling regime, as discussed below. This also causes the error bars to be larger than obtained
from other methods. Nevertheless, the critical parameters from the fit are in good agreement with the 
recently reported value of $\beta\approx1.67$.~\cite{PhysRevB.84.134209} It is also 
in general agreement with the values listed in Table~\ref{tab:critical_nuerical}. 

\begin{figure}[b]
 \includegraphics[trim = 0mm 0mm 0mm 0mm,width=1\columnwidth,clip=true]{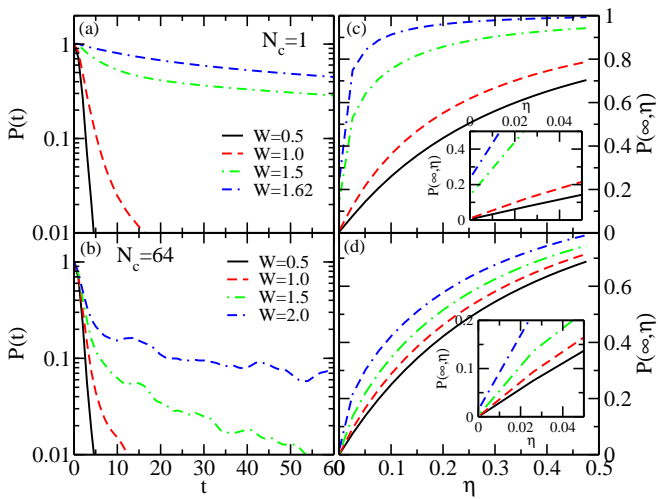}
\caption{(Color online). Right panels: The probability that an electron remains localized at 
all times $P(\infty, \eta )$ for the Anderson model for $N_c=1$ and 64 at varying disorder 
strengths.  We used the fact that 
$P(t\rightarrow\infty,\eta) = \lim\limits_{\eta \to 0} \frac{-2i\eta}{N_c} \sum_l 
\int_{-\infty}^{\infty} d\omega d\omega' \left\langle \frac{\bar{A}(l,\omega) 
\bar{A}(l,\omega)}{\omega - \omega' -2i\eta}\right\rangle$, where 
$\bar{A}(l,\omega)$ $=$ $-1/\pi \Im \bar{G}(l,l,\omega)$ is the local coarse-grained (but not 
disorder averaged) spectral function. As $\eta\rightarrow0$, $P(\infty,\eta)$ extrapolates to zero 
for small disorder strength indicating metallic behavior but does not extrapolate to zero anymore as the 
disorder strength is systematically increased towards the critical disorder strength leading to 
the transition (see the inset where this is manifestly illustrated). 
Left panels: The probability of an electron on a site thereby remaining trapped at finite 
time $t$ for $N_c=1$ and 64 for the same parameter as $P(\infty,\eta)$  on a semi-log plot.} 
\label{Fig4:pt}  
\end{figure}

Apart from the typical density of states, the localization transition in the gapless single-particle excitations of the 
Anderson insulator can be studied using the return probability of an electron to a site.~\cite{PhysRevB.5.4802,Thouless1974} The probability
of quantum diffusion (or the return probability) describes the probability of a quantum particle (or a wave) to go from 
site $l$ to $l'$ in a time $t$. After disorder averaging, the return probability has basically three key contributions: 
a) the probability of going from site $l$ to $l'$ without any scattering, b) the probability of going from site $l$ to 
$l'$ by an incoherent sequence of multiple scattering (known as diffusion) and c) the probability to go from site $l$ 
to $l'$ via a coherent (or enhanced) multiple scattering processes (e.g., the \textit{cooperon}).

In the DCA or TMDCA, it is more convenient to measure the probability of an electron remaining on a given site $l$ 
for all time $P(t\rightarrow\infty,\eta) = \lim\limits_{t \to \infty} \left\langle | G(l,l,t) |^2\right\rangle$ $
=$ $\lim\limits_{\eta \to 0} \frac{\eta}{\pi} \int_{-\infty}^{\infty} d\epsilon \left\langle | G(l,l,\epsilon+i\eta) |^2 \right\rangle$.\cite{PhysRevB.63.125102} 
This will depend on the localization length, but if a significant fraction of the eigenstates of the disordered thermodynamic
spectrum are localized states, $P(\infty,\eta)$ is expected to be nonzero.~\cite{Thouless1974,a_gonis_92,Hartmann1989,PhysRevB.63.125102} 
Since the cluster is formed by coarse-graining 
the real lattice problem in $\K$-space, there is a one-to-one correspondence between local quantities on the 
cluster and real space.~\cite{PhysRevB.63.125102} In Fig.~\ref{Fig4:pt} right panel, we show the $P(\infty,\eta)$ for the 
cluster sizes $N_c =1$ and 64 for various disorder strengths. As it is 
evident from the plot, for relatively small disorder strength $W\sim0.5$, $P(\infty,\eta)$ extrapolates to zero and 
becomes nonzero as the localization transition is approached. Just like the $P(\infty,\eta)$, the finite 
time probability that an electron on a site $l$ remains after some time $t$ denoted as: 
$P(t) = \left\langle | G(l,l,t) |^2 \right\rangle$ is a vital parameter for detecting the 
localization of electrons. As shown in Fig.~\ref{Fig4:pt} for $N_c=1$ and 64 clusters, 
$P(\infty,\eta)$ and $P(t)$ contains the same information of the 
excitation spectra. In Fig.~\ref{Fig4:pt} left panel, we show the $P(t)$ for the same parameters as $P(\infty,\eta)$. 
Hence, a characteristic finite long time $P(t)$ denotes localized eigenstates. Again, systematic 
transition from a metallic regime (for small disorder) to an insulating regime (for a disorder 
strengths close to the critical value of the Anderson localization transition in $N_c=1$ and 64 size clusters, 
respectively) is observed. Unlike in the DCA, the localization transition manifests clearly in the 
$P(\infty,\eta)$ and $P(t)$ calculated in the TMDCA since, even though the density of states (ADOS) 
calculated within the TMDCA is finite as in the DCA (the ADOS is a conserving quantity), the hybridization 
rate at the same energy depends highly on the strength of the disorder (cf. inset of Fig.~\ref{Fig2:tdos_vs_V}). 
In fact, it vanishes continuously with the disorder strength and goes to zero at the same point where 
the typical density of states vanishes. Hence, since the hybridization rate between the cluster
and the host vanishes continuously as the critical disorder strength is approached, the TMDCA 
method is able to capture the localization transition even when the ADOS calculated with the TMDCA is 
finite. 

The probability distribution function (PDF) is another natural quantity to characterize the 3D Anderson localization 
transition due to the fact that the ``typical'' value of a ``random'' variable 
corresponds to the most probable value of the PDF.~\cite{0953-8984-26-27-274209,PhysRevLett.105.046403} 
A proper description of electron localization in disordered systems requires consideration of the distribution 
functions for the quantities of interest~\cite{Anderson}, so we calculate the PDF of the cluster-momentum-resolved 
DOS $\rho(K,\omega=\bar{\epsilon}_K)$ (at different momenta cells $K$ and energy $\omega=\bar{\epsilon}_K$)
sampled over a large number of disorder configurations. 

\begin{figure}[h]
 \includegraphics[trim = 0mm 0mm 0mm 0mm,width=1\columnwidth,clip=true]{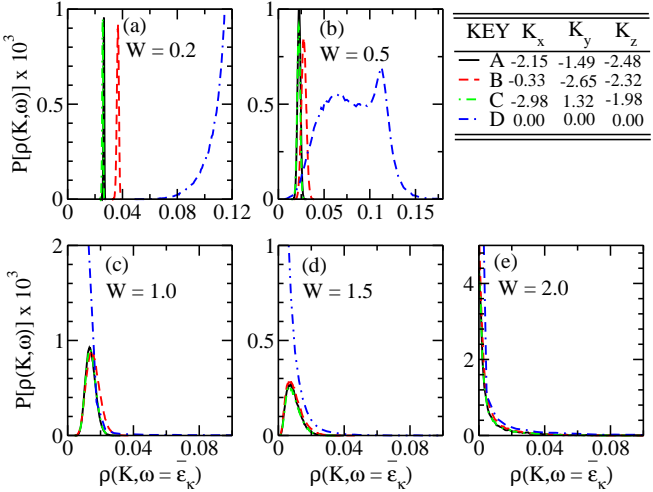}
\caption{(Color online). The evolution with disorder strength of the probability distribution of the density of states 
at different cluster cells for $N_c$ = 38. Utilizing the irreducible wedge property 
and particle-hole symmetry, the original 38 cells are reduced to 4 cells. For small disorder strength, the 
cells show Gaussian distributions whereas close to the critical disorder strength $\simeq$ 2.0, all the cells 
show log-normal distributions. }
\label{Fig3: histrogram}
\end{figure}

In Figure~\ref{Fig3: histrogram}, we show the evolution of the 
PDF[$\rho(K,\omega=\bar{\epsilon}_K)$] with $W$. As is evident from the plot, 
for a relatively small disorder, the cells show a Gaussian distribution which gradually becomes 
log-normal and highly skewed as the critical disorder strength is approached.

The analysis of the spectra properties of the Anderson model~\cite{Janssen1998,Martinelli1985} shows that for relatively 
small disorder strengths the states are still delocalized, and the amplitude of the wave functions 
associated with them is more or less the same on every site.  The distribution of the local DOS with respect 
to disorder configurations is Gaussian with the most probable value coinciding with the 
arithmetic mean average value. However, for sufficiently large disorder strength or in the proximity of the 
band tails, the spectrum consists mainly of discrete eigenvalues, and the associated eigenfunctions are 
exponentially localized with substantial weight only on a few sites. The distribution is 
therefore extremely asymmetric (log-normal), with a most probable value much 
smaller than the arithmetic mean value. At this point, most of the weight 
is concentrated around zero. As is evident from Fig.~\ref{Fig3: histrogram}, we indeed observe 
such behavior in our results.

\begin{figure}[th!]
 \includegraphics[trim = 0mm 0mm 0mm 0mm,width=1\columnwidth,clip=true]{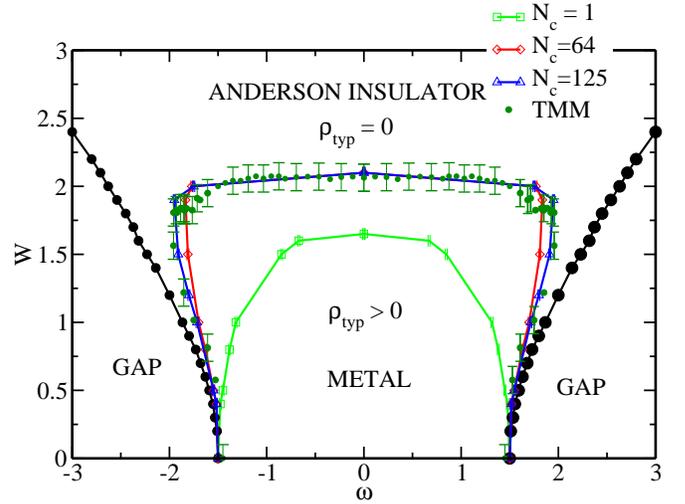}
\caption{(Color online). The phase diagram of the Anderson localization transition in 3D for the 
box disorder distribution obtained from TMDCA simulations. A systematic improvement of the 
trajectories of the mobility edge is achieved as the cluster size increases. At large enough 
$N_c$ and within computation error, our results converge to those determined by the transfer
matrix method (TMM). The TMM data is for system sizes of length 
$L = 1 \times 10^6 $ (number of multiplications), the range of system widths used is $M=[2,16]$ 
and reorthogonalization is done every $5$ transfer matrix multiplications (see Appendix \ref{TMM_appendix}
for details). The error was determined 
in a point in the region of most
discrepancy between the methods (upper bound of re-entrance) by a finite size 
scaling analysis (see Appendix \ref{TMM_appendix}). 
The black line with filled circles denotes the Lifshitz boundaries (extracted from the ADOS calculated within the DCA). 
}
\label{Fig:phaseDiagram_Box}
\end{figure}

We show in Fig.~\ref{Fig:phaseDiagram_Box}, the phase diagram of the Anderson localization transition in 
the disorder-frequency (W-$\omega$) plane constructed from our TMDCA procedure for the box disorder 
distribution.  Here, we show the mobility edge trajectories 
given by the frequencies where the TDOS vanishes at a given disorder strength $W$, and the band edge 
determined by the vanishing of the ADOS calculated within the DCA.

\begin{figure}[htb]
 \includegraphics[trim = 0mm 0mm 0mm 0mm,width=1\columnwidth,clip=true]{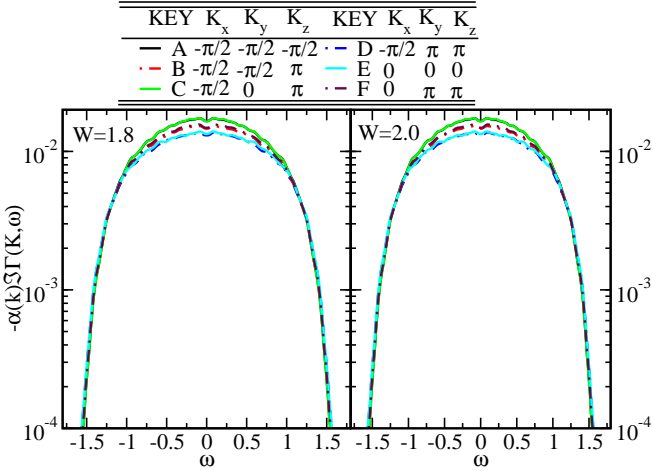}
\caption{(Color online). The scaling of the imaginary typical hybridization function ($-\Im \Gamma(\K,\omega)$) 
for the 64 site cluster at disorder strengths of W = 1.8 and 2.0 on a semi-log scale. The labels A--F and their associated momenta $\K$ 
correspond to  each of the six distinct cells obtained using the point-group and particle-hole symmetry 
($\rho(\K,\omega) = \rho(\Q-\K,-\omega)$, with $\Q= (\pi,\pi,\pi)$) of the cluster). Observe that the mobility 
edges may be collapsed on top of each other by multiplying each of the hybridization functions
with a constant such that $-\Im \Gamma(\K,\omega$) = $-\alpha(\K) \times \Im \Gamma(\K,\omega)$, where $\alpha(\K)$ 
is a scaling constant, in agreement with Mott's idea of energy selective Anderson localization transition. 
}
\label{Fig:gamma_scale}
\end{figure}

As is evident from Fig.~\ref{Fig:phaseDiagram_Box}, at $N_c$ = 64 the $W_c$ at $\omega=0$ is the 
same as that for $N_c=125$ but different from the $N_c=1$ case. This shows that the $W_c$ converges to $\sim 2.10$, 
while, the trajectory of the mobility edge continues to change with $N_c$. This may
be understood from the different localization mechanisms for states at the band center and the edge.~\cite{Bulka85,Kramer1999,Kramer1988} 
Hence, for large enough clusters, we are able to converge to the exact result.
In particular, as the cluster size $N_c$ increases, the mobility edge trajectories are systematically reproduced,
with the re-entrance behavior gradually captured with large cluster sizes. As we increase the cluster size the 
DOS systematically acquires states in the band tails, which are zero in the $N_c$=1 case
(it is well known that single-site theories  like CPA or TMT do not capture such states).
According to Bulka \etal,~\cite{Bulka85,Bulka87} deep trapped states dictate the physics at large energies. 
Hence, by making $N_c>1$ we systematically inject additional states that tend to push the localization edge outward.
States at the band center become localized mainly due to coherent backscattering while those above 
and below the bare band edges are initially localized in deeply trapped states. They become delocalized 
with increasing $W$ due to the increasing DOS at these energies and hence increasing quantum tunneling between the deeply 
trapped states. They finally become localized again with increasing disorder, which explains the 
re-entrant behavior. Since coherent backscattering requires a retracing of the electronic path,
the effective length scales captured by the cluster are doubled, such that $W_c$ converges very quickly
at the band center. On the other hand, the quantum tunneling mechanism has no path doubling and 
requires multiple deeply trapped states in the cluster and, therefore, converges more
slowly with $N_c$.

Figure~\ref{Fig:gamma_scale} is the scaling of the imaginary part of the typical hybridization function 
($-\Im \Gamma(\K,\omega)$) for $N_c=64$ at $W=1.8$ and $2.0$, scaled by the factor $\alpha(\K)$, so that 
the tails overlap. Even though different $\K$-cells go to zero at 
different rates, they each share the same unique mobility edge. Since the local disorder potential induces  
elastic scattering of a state at any momenta into any other with the same energy, there will be mixing of the localized and the 
extended states at the same energy. As a result, the localized and extended states cannot coexist at the same 
energy.  Then the mobility edge can only exist at the point where all the states in each cell on the cluster 
are localized. As shown in Fig.~\ref{Fig:gamma_scale}, the collapse of the tails for all $\K$ such that 
$-\Im \Gamma(\K,\omega$) = $-\alpha(\K) \Im \Gamma(\K,\omega)$, where $\alpha(\K)$ is a scaling constant,  
validates Mott's idea of energy selective Anderson localization.~\cite{Mott,0022-3719-20-21-008} 
 
\subsection{Alloy Model}
\label{sec:binarydistribution}
The application of the disordered tight-binding Hamiltonian (\ref{eqn:model}) to the alloy model represents one of 
the most studied physical systems. This stems from the fact that the two potentials energies W$_a$ and W$_b$ 
depict the potential landscape of e.g., a binary alloy A$_{c_a}B_{1-c_a}$, with each 
of the sites occupied either by atom ``A'' or ``B'' with concentrations $c_a$ and $c_b=1-c_a$, respectively. 

\begin{figure}[th!]
\includegraphics[trim = 0mm 0mm 0mm 0mm,width=1\columnwidth,clip=true]{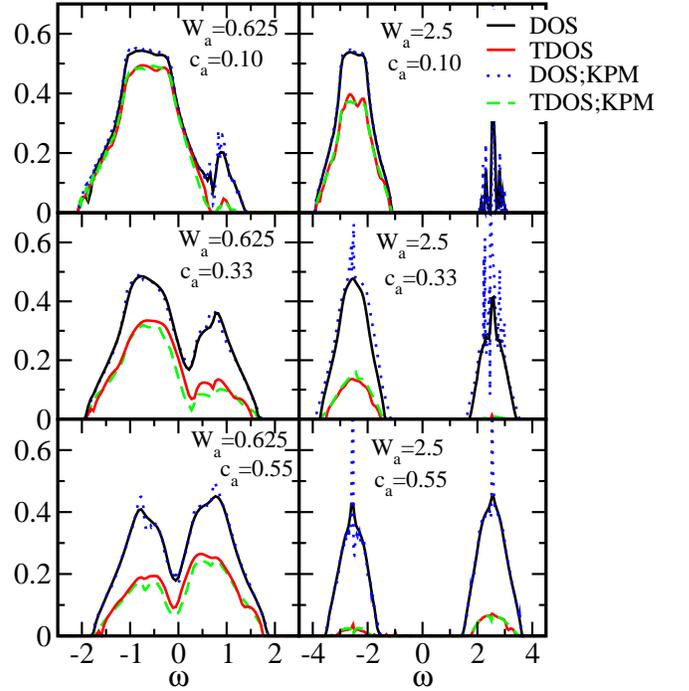}
\caption{(Color online). Comparison of the average (typical) density of states calculated 
with the DCA (TMDCA) and the kernel polynomial method (KPM) for the binary alloy system ($W_b=-W_a$)
at various values of the local potential $W_a$ and concentrations $c_a$ for the cluster size $N_c$=125. 
The kernel polynomial method uses 2048 moments on a $48^3$ cubic 
lattice, and 200 independent realizations generated with 32 sites randomly sampled for each realization.}
\label{Fig:binary_DOS}
\end{figure}


\begin{figure}[th!]
 \includegraphics[trim = 0mm 0mm 0mm 0mm,width=1\columnwidth,clip=true]{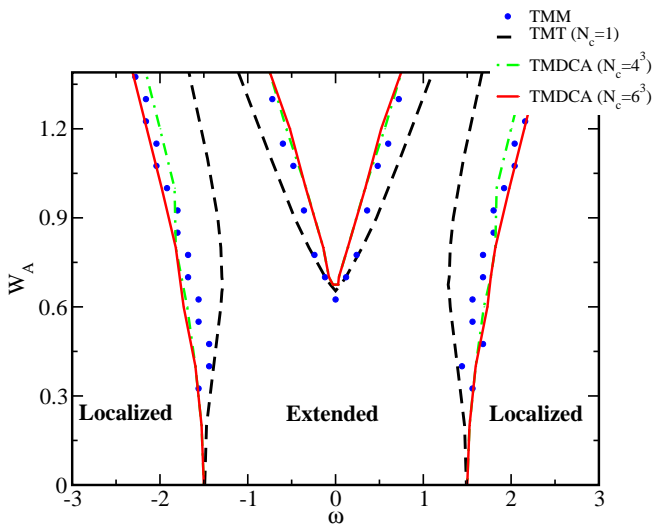}
\caption{(Color online). Disorder-energy (W$_a$-$\omega$) phase diagram 
 of the Anderson localization transition in 3D for the binary alloy system 
A$_{c_a}B_{1-c_a}$ for c$_a$ = 0.5.
Observe the systematic improvement of the trajectories of the phase diagram for the clusters ($N_c=64$ and 216) in 
basic agreement with the numerical results from TMM.  The system widths used for TMM are M=[6,12] and 
the length is scaled with the width as $L=M \times 10^4$.  See Appendix \ref{TMM_appendix} for details.  }
\label{Fig:phaseDiagram_binary}
\end{figure}

To explore the applicability of our method for the study of binary alloys, we start the discussion of this 
section by showing in Fig.~\ref{Fig:binary_DOS} the calculated typical (average) density of states from the 
TMDCA (DCA) procedure as compared to the TDOS (ADOS) calculated within the 
KPM~\cite{KPM_review_2006,Schubert,MacKinnonKramer1983,Kramer2010,Markos-review-2006} for various concentrations 
and disorder strengths. The importance of the TDOS is evident since for all 
the disorder strengths and concentrations, the ADOS remains finite around the two 
energies W$_a$ and W$_b$, while the TDOS at a fixed concentration vanishes continuously with the strength of 
the disorder with smaller values in the sub-band with the lowest concentration. Fixing the strength of 
the disorder and varying the concentration, the sub-bands with the smallest concentration have fewer states. 
We note that there are subtle differences between the results for $N_c=64$ (finite cluster) and single-site 
$N_c=1$ (CPA) (not shown) due to the incorporation of spatial correlations in the finite cluster which are 
missing in the local CPA. In fact, the TMT underestimates the extended region and misses small but important 
nonlocal features in the spectra.~\cite{PhysRevB.90.094208}

To further benchmark our results for the binary alloy model, let us focus on the 
comparison of the average (typical) DOS calculated with the DCA and TMDCA ($N_c=64$) with the KPM data.
As can be seen in Fig.~\ref{Fig:binary_DOS}, the TMDCA and DCA results reproduced those from the KPM, showing that our formalism 
offers a systematic way of studying the Anderson localization transition in binary 
alloy systems. Such a remarkable agreement is an indication of a successful benchmarking of the TMDCA 
method.~\cite{PhysRevB.89.081107} 

We extract the mobility edges shown in Fig.~\ref{Fig:phaseDiagram_binary} by finding the energy where the TDOS vanishes 
at a given value of the disorder potential. 
As can be seen in Fig.~\ref{Fig:phaseDiagram_binary}, the local $N_c=1$ boundaries are narrower than those obtained for 
the finite cluster indicating that the TMT strongly underestimates
the extended state regime. The comparison of the mobility edge boundaries for the TMDCA with those obtained from the 
TMM calculations show quite good agreement. This again is a confirmation of a successful benchmarking of the TMDCA for 
treating the binary alloy model. At the center of the band and for $c_a=0.5$, we obtain a critical disorder 
strength W$_c$ of $\approx$ 0.7 for the TMDCA in good agreement with the TMM (cf. Fig.~\ref{Fig:phaseDiagram_binary}).

\subsection{Gaussian Disorder Distribution}
\label{sec:Gaussiandistribution}
The Gaussian (or normal) (Eq.~\ref{eqn:distribution_Gaus}) is a unique distribution  
that other disorder distributions are built on
and has many physical applications including the study of molecular-doped polymers.\cite{Sin2003,Sienicki1991}

\begin{figure}[h!]
\includegraphics[trim = 0mm 0mm 0mm 0mm,width=1\columnwidth,clip=true]{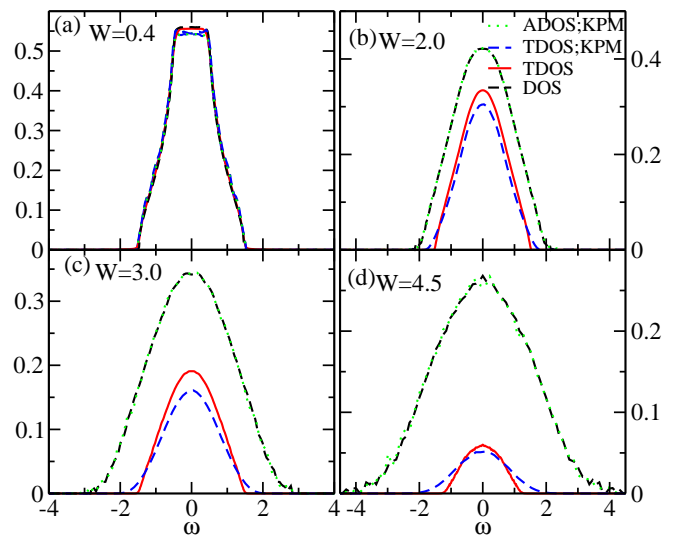}
\caption{(Color online). Comparison of the average (typical) density of states calculated with the DCA 
(TMDCA) and the kernel polynomial method (KPM) for the Gaussian disorder distribution at various values of the disorder 
strength for cluster size $N_c$=64. The kernel polynomial method uses 4096 moments on a $48^3$ cubic 
lattice, and 1000 independent realizations generated with 32 sites randomly sampled from each realization.}
\label{Fig:ADOS_TDOS_Gaussian}
\end{figure}

To further explore the versatility of our method, we apply it to study systems with 
the disorder defined by the Gaussian distribution function (Eq.~\ref{eqn:distribution_Gaus}). Again, 
we use the typical density of states (TDOS) as the order parameter and the transition to the Anderson 
insulator is obtained at the disorder strength where the TDOS vanishes. The 
typical and average density of states obtained from the TMDCA and DCA, respectively, and 
those obtained from the kernel polynomial method are shown in Fig.~\ref{Fig:ADOS_TDOS_Gaussian} 
for various values of the disorder strength. As can be seen, the TDOS at all frequencies systematically 
goes to zero as the disorder strength increases while the ADOS remains finite. Again, our TMDCA formalism 
reproduces accurately the results from the kernel polynomial method. We note some 
subtle differences between the TDOS calculated from the TMDCA and the KPM while there are 
no noticeable differences in the average density of states from the DCA and KPM. This may be due 
to the finite broadening utilized in the KPM, which contributes additional tails 
to the already exponential tails of the TDOS. We remark that aside from the small initial broadening 
value ($\sim -0.01$) used in the initialization of the TMDCA at the very first iteration, no broadening 
parameter is utilized for later iterations.

\begin{figure}[h!]
\includegraphics[trim = 0mm 0mm 0mm 0mm,width=1\columnwidth,clip=true]{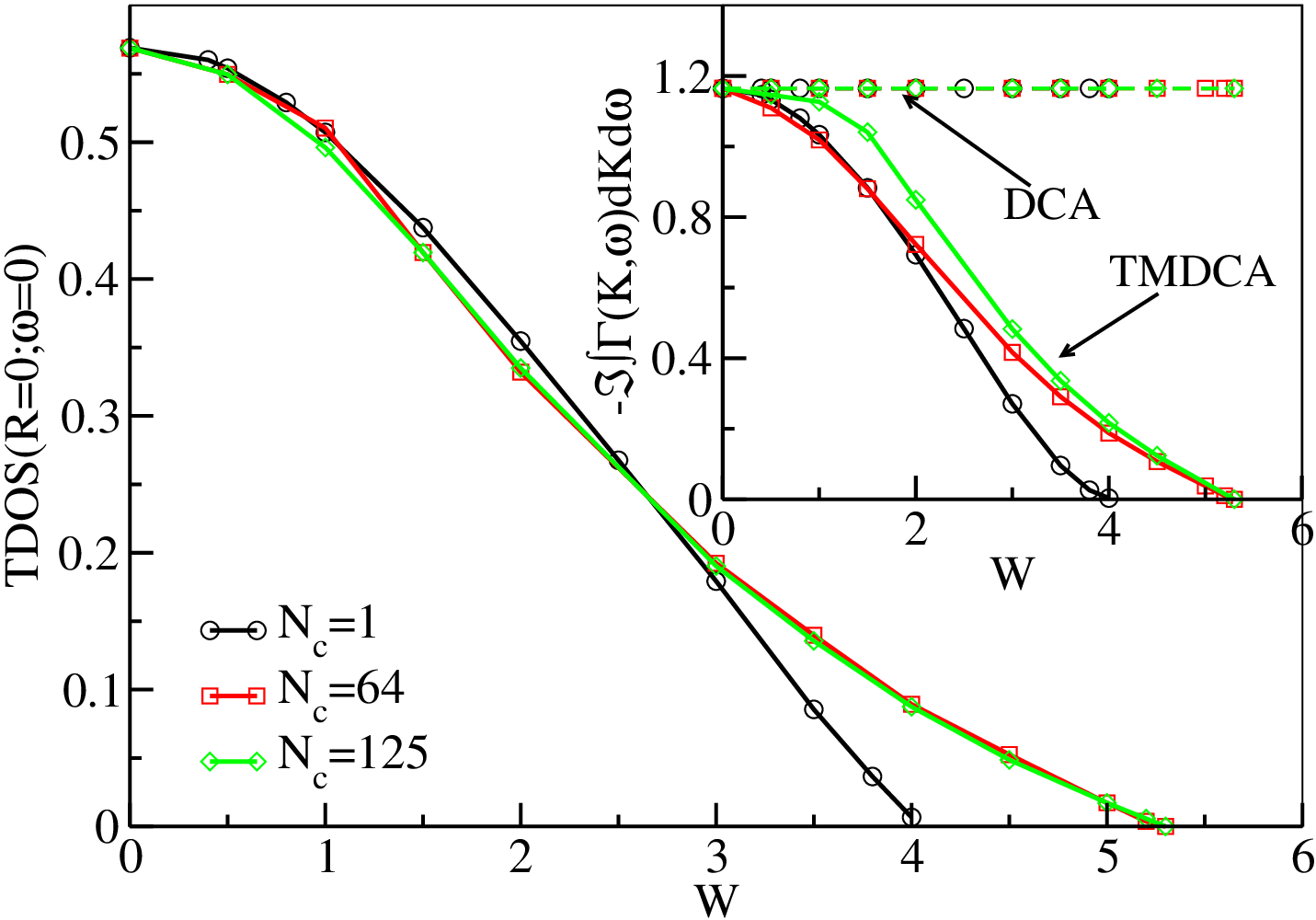}
\caption{(Color online). The TDOS$(\omega=0)$ vs.\ disorder strength $W$ at different cluster sizes $N_c=1,64,$ 
and 125 for the Gaussian disorder distribution. The TDOS$(\omega=0)$ vanishes at the critical disorder strength $W_c$.
At $N_c=1$, the critical disorder strength $W_c^{N_c=1} \approx 4.0$. 
As the cluster size increases, the critical strength systematically increases with 
$W_c^{N_c\gg12}\approx 5.30\pm 0.10$, showing a quick convergence with the cluster size. 
The inset shows 
the integrated hybridization function ($-\Im \int \Gamma(\K,\omega) d\K d\omega$) as a function of disorder 
strength $W$. Observe that $-\Im \int \Gamma(\K,\omega) d\K d\omega$ vanishes at the same disorder strength 
as the TDOS. The dashed lines are the $-\Im \int \Gamma(\K,\omega) d\K d\omega$ from the DCA. Observe that it is 
a constant regardless of the disorder strength. This shows that the DCA, even though it incorporates 
spatial correlations does not describe the Anderson localization transition. 
Moreover, near the critical region the TDOS$(\omega=0)$ data can be fitted to the 
power-law TDOS$(\omega=0)=a_0|W-W_c^{fit}|^\beta$. The obtained critical exponent for large 
enough clusters $\beta \approx 1.57\pm0.10$ is in good agreement with the numerically exact 
results.~\cite{PhysRevB.84.134209} Note, the $-\Im \int \Gamma(\K,\omega) d\K d\omega$ 
data for $N_c=64$ and 125 has been normalized with that of $N_c=1$.}
\label{Fig:TDOS_Disorder-Gaussian}
\end{figure}


We show in Fig.~\ref{Fig:TDOS_Disorder-Gaussian}, the evolution of the typical density of states TDOS($\omega=0$) 
at the band center as a function of disorder strength for the local TMT ($N_c=1$) and the TMDCA ($N_c=64$ and 125). 
Our results indicate that the critical disorder strength (defined as the $W$ where the TDOS vanishes) 
systematically increases as the cluster size is increased converging to W$_c\sim5.30$ as soon as the size of 
the cluster $N_c\ge12$.  This is in good agreement with the numerically exact values of 
5.225$\pm$0.125~\cite{Bulka87,Kramer1987,PhysRevB.51.663} and 5.32.~\cite{Slevin99} 
Fitting our data for the largest system size considered here ($N_c=125$) using the power law: 
TDOS$(\omega=0)=a_0|W-W_c^{fit}|^\beta$ and following the procedure as explained in 
Subsection~\ref{criticalparameters}, we obtain the order parameter critical exponent $\beta\sim$ $1.57\pm0.10$ with 
a corresponding critical disorder strength from the fit of $W_c^{fit}\sim5.53\pm0.10$. This value of $\beta$ 
is in good agreement with the value we obtained for the uniform disorder 
distribution (cf. Table~\ref{Tab:critical_nu}) and in good agreement with the recently reported value 
of $\beta\approx1.67$.~\cite{PhysRevB.84.134209} It is also 
in general agreement with the values listed in Table~\ref{tab:critical_nuerical}. The good agreement between the $\beta$ 
we obtained from the uniform and Gaussian disorder distribution is a manifestation of the universal nature 
of the Anderson localization transition.~\cite{PhysRevB.51.663,Wegner1989}    

\begin{figure}[th!]
 \includegraphics[trim = 0mm 0mm 0mm 0mm,width=1\columnwidth,clip=true]{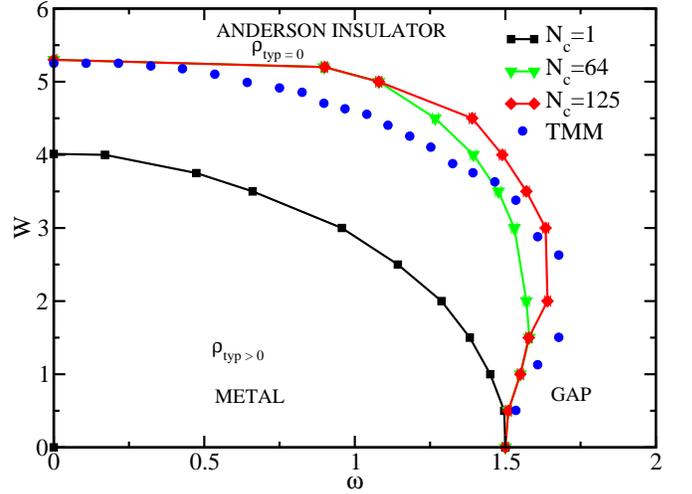}
\caption{(Color online). The phase diagram of the Anderson transition in 3D for the Gaussian disorder distribution 
obtained from TMDCA simulations. A systematic improvement of the trajectories of the mobility edge is achieved as the 
cluster size increases. At large enough $N_c$, our results 
converge to those determined by the TMM which are calculated for widths $M=[2,16]$ and for
a system length (number of transfer matrix multiplications)  $L = 1 \times 10^6$, where the 
matrix products are reorthogonalized every $5$ transfer
matrix multiplications (see Appendix \ref{TMM_appendix}). The deviation between the TMDCA and TMM results is consistent with
the behavior observed for the box and Lorentzian disorders and can be attributed to the fact that a finite grid in energy
is used for the TMDCA which tends to cause the typical density of states to be larger, hence slightly
overemphasizing the metallic behavior and as such, the mobility edge is slightly larger 
when compared to TMM in certain frequency ranges near the band edge. The effect is most pronounced
here due to the small density of states near the band edge. In addition, near the re-entrance
regime, the TMM also has difficulties due to an increase in finite size effects (see Appendix \ref{TMM_appendix}).}

\label{Fig:phasediagram_Gussian}
\end{figure}

To explore the trajectories of the mobility edge for the Gaussian disorder distribution, we show in 
Fig.~\ref{Fig:phasediagram_Gussian} the phase diagram in the energy-disorder plane for various cluster 
sizes as compared to the TMM result.~\cite{Bulka87} For any given 
disorder strength $W$, the mobility edge is defined by the frequency where the TDOS vanishes. 
Unlike the critical disorder strength which converges quickly 
with the cluster size $N_c\geq$ 12, the trajectory of the mobility edge continues to change 
with $N_c$ converging to almost the numerically exact results for $N_c=125$. The physical 
reasons for the quick convergence of W${}_c$ and the progressive change of the mobility edge with 
the cluster size are the same as those described above for the box disorder 
distribution (cf. Section~\ref{sec:boxdistribution}). As can be seen, the single-site 
TMT underestimates the extended region just as in the previously presented disorder distributions. There 
are, however, some subtle differences between our data and the TMM results around the re-entrance regime. 
The cause of this difference will be discussed in Section~\ref{sec:difficultyinME}.

\subsection{Lorentzian Disorder Distribution}
\label{sec:Lorentziandistribution}
\begin{figure}[t]
\includegraphics[trim = 0mm 0mm 0mm 0mm,width=1\columnwidth,clip=true]{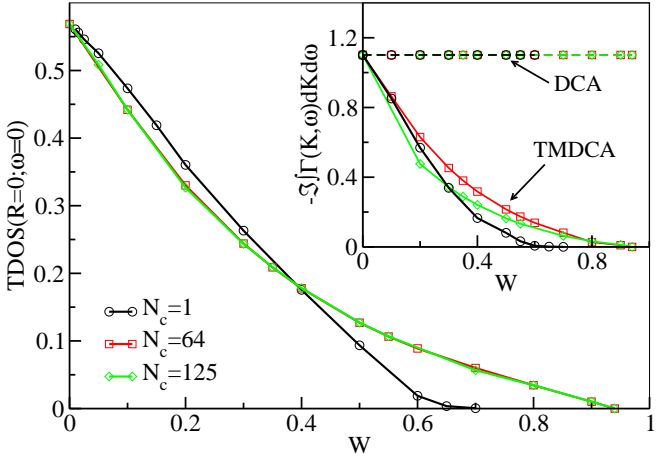}
\caption{(Color online). The TDOS$(\omega=0)$ vs.\ disorder strength $W$ at cluster sizes 
$N_c=1$, $64$, and $125$ for the Lorentzian disorder distribution. The TDOS$(\omega=0)$ vanishes at 
the critical disorder strength $W_c$.  At $N_c=1$, the critical disorder strength $W_c^{N_c=1} \approx 0.6$. 
As the cluster size increases, the critical strength systematically increases until 
$W_c^{N_c\gg12}\approx 0.94\pm 0.10$, showing a quick convergence with the cluster size. 
The inset shows 
the integrated hybridization function ($-\Im \int \Gamma(\K,\omega) d\K d\omega$) as a function of disorder 
strength $W$. Observe that $-\Im \int \Gamma(\K,\omega) d\K d\omega$ vanishes at the same disorder strength 
as the TDOS. The dashed lines are the $-\Im \int \Gamma(\K,\omega) d\K d\omega$ from the DCA. Observe that it is 
a constant regardless of the disorder strength in agreement with the observations we have for other 
disorder distributions. We can fit the TDOS$(\omega=0)$ data near the critical region to the 
power-law, with TDOS$(\omega=0)=a_0|W-W_c^{fit}|^\beta$. The obtained critical exponent for large 
enough clusters $\beta \approx 1.60 \pm 0.10$ is in good agreement with numerically exact 
results.~\cite{PhysRevB.84.134209} Note, the $-\Im \int \Gamma(\K,\omega) d\K d\omega$ 
data for $N_c=64$ and 125 have been normalized with that of $N_c=1$.}
\label{Fig:TDOS_Disorder-LD}
\end{figure}

We next apply our TMDCA formalism to study systems with the Lorentzian (or Cauchy) disorder 
distribution (cf. Eq.~\ref{eqn:distribution_Lo}). 
We show in Fig.~\ref{Fig:TDOS_Disorder-LD} how the band center of the typical density 
of states (TDOS($\omega=0$)) changes as the disorder strength is increased for the local TMT ($N_c=1$) 
and the TMDCA ($N_c=64$ and 125). As can be seen from Fig.~\ref{Fig:TDOS_Disorder-LD}, our results depict that the 
critical disorder strength systematically increases as the cluster size increases converging to W$_c\sim0.94$ for 
$N_c\ge12$ in good agreement with the numerically exact values of 0.95$\pm$0.125~\cite{Bulka87,Kramer1987,PhysRevB.51.663} 
and 1.07,~\cite{Slevin99} respectively. 
Fitting our data for the largest system size considered here ($N_c=125$) using the power law: 
TDOS$(\omega=0)=a_0|W-W_c^{fit}|^\beta$ (see Subsection~\ref{criticalparameters} for the 
description of how the $\beta$ is extracted), we can infer the order parameter critical 
exponent $\beta\sim$ 1.60$\pm$0.10 with 
a corresponding critical disorder strength from the fit, $W_c^{fit}\sim0.97\pm0.10$. The obtained $\beta$ from 
our fit is in good agreement with the values listed in Table~\ref{tab:critical_nuerical} 
and in good agreement with the recently reported value of $\beta\approx1.67$.~\cite{PhysRevB.84.134209} This 
$\beta$ value of the Lorentzian disorder distribution is also in good agreement with the values 
we obtained for the box and Gaussian disorder distributions, respectively. This further illustrates the universal 
nature of the Anderson localization transition.~\cite{PhysRevB.51.663,Wegner1989}

\begin{figure}[th]
 \includegraphics[trim = 0mm 0mm 0mm 0mm,width=1\columnwidth,clip=true]{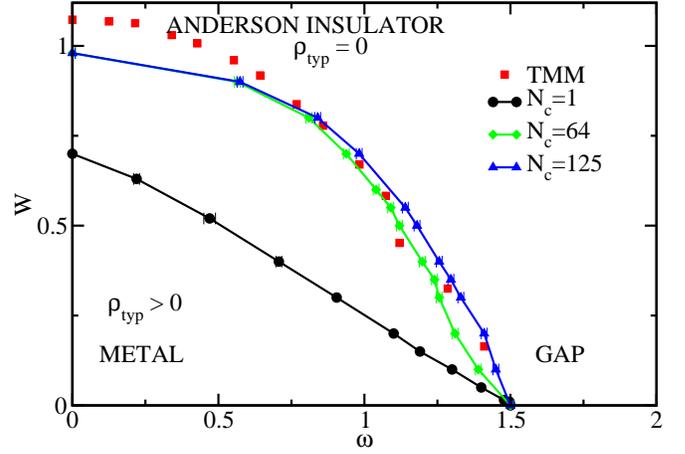}
\caption{(Color online). The phase diagram of the Anderson localization transition in 3D for the Lorentzian disorder 
distribution obtained from TMDCA simulations. 
The TMM data were computed for a system length of $L=8 \times 10^5$
for system widths of $M=[ 2,16]$ and re-orthogonalization is done every four multiplications (see Appendix \ref{TMM_appendix} 
for details). The discrepancy between the TMDCA and TMM is due to the nature of the Lorentzian distribution 
as explained in Sec.~\ref{sec:difficultyinME}.}
\label{Fig:phasediagram_LD}
\end{figure}

We conclude our study of the application of the typical medium dynamical cluster approximation 
to the Lorentzian disorder distribution by presenting in Fig.~\ref{Fig:phasediagram_LD} the phase 
diagram in the energy-disorder plane. Unlike the box and Gaussian disorder distributions, our simulations 
show that the Lorentzian distribution does not have re-entrance of the mobility edge. The lack of 
re-entrance of the mobility edge in the Lorentzian disorder distribution may be 
attributed to the absence of finite variance in this form of distribution. For the single-site
CPA (N$ {} _c=1$), the critical parameters are woefully underestimated. However, we systematically converge 
to the numerically exact results as the size of the cluster is increased.
As it is obvious from Fig.~\ref{Fig:phasediagram_LD}, for as small as N$ {} _c=64$, we converge 
almost to the exact TMM results.~\cite{Bulka85} We again see remarkably good 
agreement between our effective mean-field method for the Anderson localization transition and 
the numerically exact calculations.   

\subsection{Critical Parameters}
\label{criticalparameters}
The critical parameters, including the critical disorder strength $W_c$ and the order parameter 
critical exponent $\beta$ are summarized in Table~\ref{Tab:critical_nu} for different cluster sizes 
using the box disorder distribution as a case study. $W_c^{cal}$ was determined as
the $W$ where the TDOS\,($\omega=0$) vanishes. Observe that as $N_c$ increases, $W_c^{cal}$ systematically 
increases with $W_c^{cal}\approx 2.10 \pm 0.01$, showing a quick convergence with $N_c$. 
\begin{table}[thb] 
\centering
\caption{The calculated and fitted critical disorder strengths $W_c^{cal}$ and $W_c^{fit}$ and the 
order parameter critical exponent $\beta$ obtained from our fit for the box disorder 
distribution.  $W_c^{cal}$ is defined by the vanishing of the TDOS\,($\omega=0$). 
$\beta$ and $W_c^{fit}$ are obtained by fitting the 
TDOS$(\omega=0)$ data with a power law, TDOS$(\omega=0)=a_0|W-W_c^{fit}|^\beta$.}
\begin{tabular}{cccccccc}
\hline\hline
$N_c$ & W$_c^{cal}$ & W$_c^{fit}$ & $\beta$ \\ 
\hline
1 & 1.66$\pm$0.01 & 1.65$\pm$0.10 & 0.96$\pm$0.10 \\ 
64 & 2.10$\pm$0.01 & 2.18$\pm$0.10 & 1.46$\pm$0.10 \\ 
125 & 2.10$\pm$0.01 & 2.23$\pm$0.10 & 1.62$\pm$0.10 \\ 
\hline\hline
\end{tabular}
\label{Tab:critical_nu}
\end{table}

\begin{figure}[b!]
\includegraphics[trim = 0mm 0mm 0mm 0mm,width=1\columnwidth,clip=true]{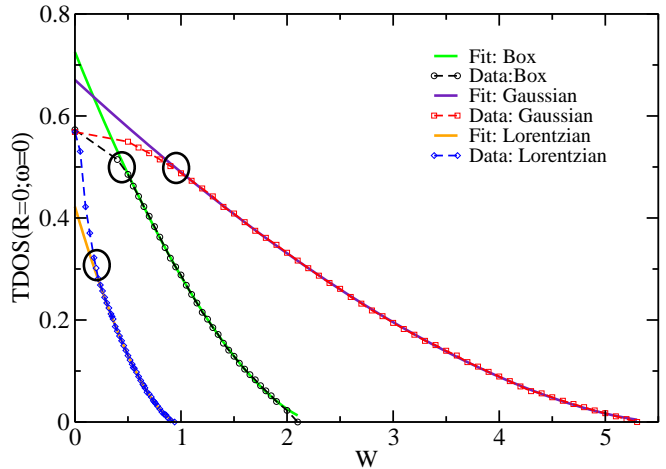}
\caption{(Color online). The band center of the typical density of states (TDOS($R=0; \omega=0$)) at 
various disorder strengths ($W$) for the 125 site cluster for the 
Box, Gaussian, and Lorentzian disorder distributions. The linear region of the 
data can be fit with a scaling ansatz: TDOS$(\omega=0)=a_0\times|W-W_c^{fit}|^\beta$, 
with $\beta\sim1.62$ for the Box disorder distribution, $\beta\sim1.57$
for the Gaussian disorder distribution, and $\beta\sim1.60$ for the Lorentzian disorder distribution
in good agreement with the recently reported 
value of $\beta\approx1.67$.~\cite{PhysRevLett.105.046403,PhysRevB.84.134209} The circles in the plot 
depict the data point where the fit starts to deviate from the data.}
\label{Fig:scaling_fit}
\end{figure}

The order parameter critical exponent $\beta$ is obtained by fitting the power law: 
TDOS$(\omega=0)=a_0\times|W-W_c^{fit}|^\beta$ directly to our data and systematically 
searching for the best data point away from the transition where the fit still follows 
the actual data (scaling regime). This becomes imperative since away from the transition, 
the data are not expected to fit the form TDOS$(\omega=0)=a_0\times|W-W_c^{fit}|^\beta$ and 
also close to the transition, there should be a crossover to a mean-field form since the TMDCA treats 
the longest length scales in a mean-field approximation.   
So, the fit may only be done between these limits. The ambiguity in the determination
of the fitting region increases our error bars on the exponent, and causes $W_c^{fit}$ to
be overestimated.  In addition, the strong fluctuation of the TDOS in the proximity of the 
critical point also increases our error bars. We show in Fig.~\ref{Fig:scaling_fit} 
a comparison plot of the fit and our data for the 125 site cluster for the box, Lorentzian, and 
Gaussian disorder distributions, respectively. The fit of the power law to the scaling region of 
the data gives the value of the $\beta$ in an unambiguous manner. The obtained values of $\beta$ 
from the various cluster sizes, for instance, for the box disorder distribution are shown in 
Table~\ref{Tab:critical_nu}. One can see that our $\beta$ systematically approaches the 
numerical experimental value~\cite{Slevin99,PhysRevLett.105.046403,PhysRevB.84.134209} 
for large enough clusters (here, largest $N_c$ simulated is 125) as listed in 
Table~\ref{tab:critical_nuerical} for the box, Gaussian, and Lorentzian disorder distributions, 
respectively, and in comparison with other numerical values. 

\begin{table*}[htb]
\centering
\caption{Critical parameters of the Anderson localization 
for the various studied disorder distributions in 3D obtained using the TMDCA in comparison with the numerically exact results. 
We use $4t=1$ as our energy unit. We note that the critical exponents ($\beta$ and $\nu$) 
are independent of disorder distribution (universal) as verified by the multifractal analysis,~\cite{PhysRevB.51.663} the analytic 
results of Wegner,~\cite{Wegner1989} and detailed finite size scaling.~\cite{Slevin2014}  
Abbreviations used in the Table are: transfer matrix method (TMM), multifractal finite size scaling
(MFSS), level statistics (LS), kicked rotor (KR), and experimental atomic waves (Exp-AW).} 
\begin{tabular}{ccccccc}
\hline\hline
Author & \multicolumn{3}{c}{Critical Disorder} & \multicolumn{2}{c}{Critical Exponent} & Method \\ 
\cline{2-4} \cline{5-6}
 & Bo & Ga & Lo & $\nu$ & $\beta$ &  \\ 
\hline 
Present study & 2.10$\pm$0.10& 5.30$\pm$0.10& 0.94$\pm$0.10 & --&1.57--1.62 & TMDCA \\ 
Slevin \etal \footnotemark[4] & 2.067& --& 1.067 & 1.573--1.577&-- & TMM \\ 
Slevin \etal \footnotemark[5] &2.068--2.073 & 5.32& 1.066 & 1.58&-- & TMM \\ 
Slevin \etal \footnotemark[6] &2.056$\pm$0.014 & --& -- & 1.59--1.60&-- & TMM \\ 
Rodriguez \footnotemark[7] &2.066--2.067 & --& -- &1.59& 1.65--1.68 & MFSS \\ 
Rodriguez \footnotemark[8] &2.066--2.071 &-- & -- & 1:58$\pm$0.03& -- & MFSS \\ 
Markos\footnotemark[9] & 2.063,2.067 &-- & --& 1.47--1.55& --& MFSS \\ 
MacKinnon\footnotemark[10] &2.063 $\pm$0.05 &-- & -- & 1.54$\pm$0.08 & --& TMM  \\ 
MacKinnon \etal \footnotemark[11] &2.063 $\pm$0.063&--& -- & 1.2$\pm$0.3& --& TMM  \\ 
Bulka \etal \footnotemark[12] & 2.038--2.063& 5.23$\pm$0.13& 0.95$\pm$0.13 & -- & --& TMM \\ 
Milde \etal \footnotemark[13] & --& --& -- & 1.62$\pm$0:07 &-- & TMM \\
Shklovskii \etal \footnotemark[14] &2.0$\pm$0.063&--& -- & 1.50$\pm$0.15 & & LS \\ 
Zharekeshev \etal \footnotemark[15] &2.05&--& -- & 1.4$\pm$0.15 &-- & LS \\ 
Hofstetter \footnotemark[16] & 2.719$\pm$0.012& --& -- & 1.35$\pm$0.15 &-- & LS \\ 
Lopez \footnotemark[17] &-- & --& --& 1.63$\pm$0.05 &--& KR\\ 
Grussbach \etal \footnotemark[18] &2.02 & 5.23& -- & 1.32$\pm$0.02--1.37$\pm$0.02 &1.32$\pm$0.02--1.37$\pm$0.02& MFSS
\footnote{The authors obtain $\alpha_o=4.0$ such that in the hyperscaling relation $\beta = (\alpha_o-d)\nu$, 
$\beta$ equals $\nu$.}   \\ 
Lemari\'e \footnotemark[19] & --& --& -- & 1.58$\pm$0.01--1.60$\pm$0.03 &--& KR \footnote{Authors of this 
paper show that their quasiperiodic kicked rotor belongs to the same (orthogonal) universality class as the 
`random' Anderson model.} \\ 
Lemari\'e \footnotemark[19] &-- & --& -- & 1.40$\pm$0.30 &--& Exp-AW \footnote{Authors of this 
paper reported that experiments were done on the atomic kicked rotor by a sequence of 
kicks to the atomic cloud and measure its dynamics.} \\ 
\hline\hline
\end{tabular}
\begin{tablenotes}
 \small 
\item The order parameter critical exponent $\beta$ and the correlation length critical exponent $\nu$ 
can be transformed from one to the other using the hyperscaling relation of Ref.~(\onlinecite{PhysRevB.84.134209}) 
$\beta = (\alpha_o-d)\nu$, where $\alpha_o$ is the Lipschitz-H\"{o}lder exponent, which gives the maximum value of the 
multifractal spectrum. 
The most recent estimates as reported in Ref.~(\onlinecite{PhysRevB.84.134209}) are 
$\alpha_o = 4.048$, $\nu=1.59$ and $\beta=1.67$.
\end{tablenotes}
\footnotetext[4]{Ref.~\onlinecite{Slevin2014}.}
\footnotetext[5]{Ref.~\onlinecite{Slevin99,PhysRevB.63.045108}.}
\footnotetext[6]{Ref.~\onlinecite{PhysRevLett.78.4083}.}
\footnotetext[7]{Ref.~\onlinecite{PhysRevB.84.134209}.}
\footnotetext[8]{Ref.~\onlinecite{PhysRevLett.105.046403}.}
\footnotetext[9]{Ref.~\onlinecite{0305-4470-33-42-103}.}
\footnotetext[10]{Ref.~\onlinecite{0953-8984-6-13-012}.}
\footnotetext[11]{Ref.~\onlinecite{PhysRevLett.47.1546}.}
\footnotetext[12]{Ref.~\onlinecite{Bulka87,Bulka85}.}
\footnotetext[13]{Ref.~\onlinecite{Milde2000}.}
\footnotetext[14]{Ref.~\onlinecite{PhysRevB.47.11487}.}
\footnotetext[15]{Ref.~\onlinecite{PhysRevLett.79.717}.}
\footnotetext[16]{Ref.~\onlinecite{PhysRevB.57.12763}.}
\footnotetext[17]{Ref.~\onlinecite{PhysRevLett.108.095701}.}
\footnotetext[18]{Ref.~\onlinecite{PhysRevB.51.663}.}
\footnotetext[19]{Ref.~\onlinecite{PhysRevA.80.043626,0295-5075-87-3-37007}.}
\label{tab:critical_nuerical}
\end{table*}

\subsection{Difficulties in Extracting the Mobility Edge at Higher Disorder}
\label{sec:difficultyinME}
As explained in the previous sections, the mobility edge is obtained by locating 
the frequencies where the typical density of states vanishes at a given disorder strength $W$. The band edge is
determined by the vanishing of the average density of states calculated within the DCA. As it is obvious from the 
phase diagrams for the various disorder distributions (Figs.~\ref{Fig:phaseDiagram_Box}, \ref{Fig:phasediagram_Gussian}, 
and \ref{Fig:phasediagram_LD}, respectively), there are some discrepancies between the phase diagram obtained 
within the typical medium dynamical cluster approximation and the transfer matrix method. They can 
be attributed in part to the form of the disorder distribution. For example, the bare DOS for the 
Gaussian and Lorentzian disorder distributions are known to have exponential tails. 
The severity of the exponential nature of the tails associated with the various disorder distributions increases as box, 
Gaussian, and Lorentzian, in that order. At these higher disorder strengths (once a mobility edge develops), 
the TDOS naturally develops tails. 
\begin{figure}[b!]
 \includegraphics[trim = 0mm 0mm 0mm 0mm,width=1\columnwidth,clip=true]{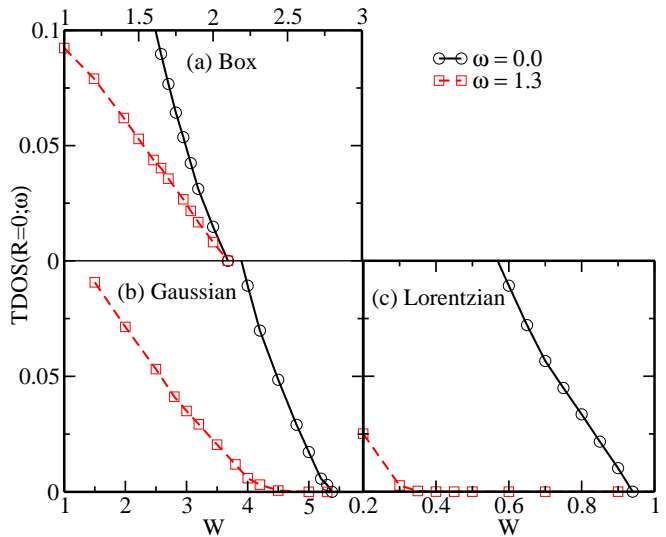}
\caption{
(Color online). The plot of the typical density of states (TDOS) at $\omega=0.0$ and $\omega=1.3$ for
various disorder strengths for the (a) Box, (b) Gaussian, and (c) Lorentzian disorder distributions. }
\label{Fig:compare_ws}
\end{figure}
We note that in our computations, apart from the initial small broadening value $\sim -0.01$ used in the 
initialization of the self-energy (needed only for the first iteration), no broadening factor is utilized. 
As such, these tails that emerge as the disorder strength is increased towards $W_c$ are physical tails since 
the top and bottom of the bands will localize first. To demonstrate this, 
we show in Fig.~\ref{Fig:compare_ws} a plot of the TDOS at $\omega=0.0$ and $\omega=1.3$
for the box, Gaussian, and Lorentzian disorder distributions, respectively. 
Note the $\omega=1.3$ frequency is arbitrary but chosen  such that it is close to the 
re-entrance region of the mobility edge.
As can be seen from the plots, for the box disorder distribution, even though there are small tails, the TDOS at 
$\omega=0.0$ and $\omega=1.3$ behave alike and differ only in magnitude. There are no obvious long 
tails in either frequency ($\omega=0.0$ and 1.3) that may mask the detection of the position of 
the mobility edge energies. However, for the Gaussian and 
Lorentzian disorder distributions, there are exponentially long tails especially at 
$\omega=1.3$ that make pinpointing the 
exact position of the mobility edge energies highly non-trivial. This should further 
be understood from the fact that unlike the box disorder 
distribution, the Lorentzian disorder distribution naturally has tails that decay 
very slowly at infinity, as $|x|^{-2}$, such that aside the zeroth moment 
(the area under the curve), all other higher moments do not even exists.

We note that this difficulty is generic not only to the TMDCA but also for 
any method where the extraction of the trajectories of the mobility edge is based on the TDOS. 
A notable example is in the kernel polynomial method, which even shows more severe discrepancies (not shown). 

This deviation between the TMDCA and TMM at higher disorder strengths can also be attributed 
to the fact that the TMDCA, just like the KPM utilizes a finite frequency grid (which biases more 
towards the metallic regime) in contrast to the TMM which calculates the transmission of 
electrons at fixed energies. Even so, we note that the TMM also has its own 
shortcomings away from the band center due to the strong fluctuations in the 
Kramer-MacKinnon scaling parameter $\Lambda$ (see Appendix~\ref{TMM_appendix} for its definition) as can 
be seen from the phase diagrams for the various disorder strengths around the 
re-entrance regime.


\section{Conclusions}
\label{sec:conclusion}
Here, we present a detailed study of the Anderson localization transition using the recently developed typical 
medium dynamical cluster approximation (TMDCA) for the box, Gaussian, Lorentzian, and binary alloy disorder 
distributions in three dimensions. For each distribution we find the TMDCA to be a successful, causal, numerically 
efficient, self-consistent and rapidly convergent method for the study of localization in disordered electron systems. 

With our formalism, we demonstrate that the typical DOS vanishes for localized states and is finite for extended 
ones. Employing the typical DOS as an order parameter for Anderson localization, we have constructed the 
disorder-energy phase diagram,  extracted the order parameter critical exponent ($\beta$) for each disorder 
distribution, and benchmarked them in good agreement with other numerically exact methods. Within our precision, we 
find that $\beta$ for the Anderson localization transition is a universal parameter independent of disorder 
distribution in  agreement with the multifractal analysis.~\cite{PhysRevB.51.663} For distributions with a finite 
variance (box and Gaussian), we demonstrate that there are extended states outside the unperturbed band.

We further show using the DCA (which includes spatial correlations) and a variant of the typical medium theory
(which includes spatial correlations but suffers from self-averaging), the importance of the effective medium to 
properly characterize the Anderson localization transition. We also demonstrate the inability of the single-site 
CPA and the TMT methods to accurately capture the localization and disorder effects in both the average and the typical 
DOS. We note that the single-site TMT, while being able to qualitatively capture the localization transition, 
strongly underestimates the extended regions and fails to capture the critical parameters including the mobility edge 
trajectories and the exponents. In contrast, the TMDCA captures nicely the trajectories of the mobility edge with great 
improvement in the critical order parameter exponent. Most significantly, the TMDCA results are in a quantitative 
agreement with exact numerical results.

The TMDCA formalism is computationally inexpensive and straightforward to implement since it requires only the 
computer time needed to invert small clusters (e.g., $N_c=1$--125), average over the disorder configurations, 
and iterate to convergence.  Since only a small cluster is needed to get reliable data, material specific details 
may be incorporated. Once combined with electronic structure calculations\cite{PhysRevB.41.9701} and more 
sophisticated many-body techniques for electron interactions, it will open a new avenue for studying the localization 
phenomenon in real materials as well as the competition between disorder and electron correlations. To demonstrate 
the high efficiency of the TMDCA, as only small clusters are needed to get a converged result in good agreement 
with the TMM data, we compare the relative CPU time needed for the largest system size simulated in the TMDCA 
and the TMM.  For the largest cluster size used in the TMDCA calculations, which is $N_c=216$, the computation 
time is $\sim$4 hours (running on a single processor).  While for the TMM, which is perfectly parallel in both disorder and 
frequency, each point in the phase diagram can require significant computational effort.  For example, the system 
sizes in Fig.~\ref{Fig:phaseDiagram_Box} used $\sim$20 hours on 64 processors per frequency.  Since a separate TMM 
calculation is needed for each frequency, achieving the energy resolution of a typical TMDCA calculation (for a 
certain number of frequency grid points) would require the product of the number of processors used to parallelize  
over disorder times the number of grid points.  As a calculation of a real material would require even larger 
system sizes than used here in the TMM, the TMDCA would prove much more computationally efficient for the purpose of 
studying real materials.

\textit{Acknowledgments}--
We thank  Shuxiang Yang and Vladimir Dobrosavljevi\'{c} for useful discussions. 
We also thank Janee Jarrell for a careful reading of the manuscript.
This work is supported by NSF DMR-1237565 and NSF EPSCoR Cooperative 
Agreement EPS-1003897 with additional support from the Louisiana Board of 
Regents, and the DOE Computational Materials and Chemical 
Sciences Network (CMCSN) SC0007091. Supercomputer support was provided by the Louisiana Optical 
Network Initiative(LONI), and HPC@LSU computing resources.
\appendix
\section{Details of the Transfer Matrix Method}
\label{TMM_appendix}
We benchmark our TMDCA results for the mobility edge with the transfer matrix method 
(TMM),~\cite{PhysRevLett.47.1546,Kramer,Kramer1996,MacKinnonKramer1983,Kramer2010,Markos-review-2006} 
which is an established numerical method for determining the mobility edge by computing the localization length 
in disordered quantum systems. TMM is based on an iterative formulation of the Schr\"{o}dinger equation
where the wave function amplitude is computed at each site in a quasi-one-dimensional ``bar'' of length 
$L$ and width $M$ by successive multiplications of the transfer matrix that describes the transmission 
between each ``slice'' of size $M^{d-1}$.  Thus, the Lyapunov exponent that measures the exponential 
decay of the wavefunction is explicitly computed, yielding the localization length. The system length
$L$ is the total number of transfer matrix multiplications. The numerical instability of the repeated 
multiplications is avoided by periodically orthogonalizing the transfer matrix product with a Lapack 
QR decomposition after a finite number of iterations.~\cite{Slevin2014}  The transfer matrix method 
finite size effects are larger for weak disorder where the states decay slowly with distance and so 
have large values of the localization length, which results in more pronounced fluctuations in the data. Notice that 
the CPA and the DCA (same as the TMDCA) do not suffer such finite size effects for small disorder and are, in fact, exact 
in this limit.

The mobility edge is obtained by calculating the dimensionless Kramer-MacKinnon scaling 
parameter $\Lambda_M$, which is the localization length divided by $M$.~\cite{MacKinnonKramer1983} 
$\Lambda_M$ scales as a constant for $M \rightarrow \infty$ at the transition.~\cite{Kramer2010}  
Precise values of the critical 
disorder may be measured directly from the crossing plots of $\Lambda_M$ vs. $W$. 
A finite size scaling is performed by expanding $\Lambda_M$ near the critical point using
\begin{equation}
\Lambda_M = f(M/ \xi) \approx \Lambda_c + a_1 |W-W_c| M^{1/\nu} + a_2 |W-W_c|^2 M^{2/\nu} + \cdots
\end{equation} 
and the data is fit using a least squares procedure.~\cite{nakayama2003fractal} 
The data used in Fig.~\ref{Fig:phaseDiagram_Box} was a third order polynomial in $|W-W_c| M^{1/\nu}$.  
The critical disorder strength $W_c$ from the finite size scaling
is averaged over many generated data sets via a bootstrap procedure.~\cite{Slevin2014} 
Any errors quoted in the TMM data are from the difference in the measured critical disorder from the finite size 
scaling analysis\cite{MacKinnonKramer1983} and the critical disorder measured directly from the crossing
plots of $\Lambda_M$ vs. $W$.

\section{Details of the Kernel Polynomial Method}
\label{KPM_appendix}
To further benchmark our results, we utilize the kernel polynomial
method to calculate the local DOS.~\cite{SILVER1994,PhysRevE.56.4822,KPM_review_2006, Schubert}
In the kernel polynomial method analysis, instead of diagonalizing the Hamiltonian directly,
the local DOS is expanded in terms of an infinite series of Chebyshev polynomials. In practice,
the truncated series leads to Gibbs oscillations. The kernel polynomial method damps these
oscillations by a modification of the expansion coefficients. 
We use the Jackson kernel following previous studies on the Anderson model.~\cite{KPM_review_2006}


\end{document}